\documentclass[aps,prb,showpacs,twocolumn,floats,epsfig,pdflatex]{revtex4}
\usepackage{amssymb}
\usepackage{amsbsy}
\usepackage{amsmath}
\usepackage{epsfig}
\usepackage{amsfonts}
\usepackage{graphicx}
\usepackage{amssymb}
\usepackage{upgreek}
\usepackage{dsfont}
\usepackage{bm}
\usepackage{color}
\usepackage{hyperref}

\def\grad{\boldsymbol\nabla}

\def \E{|{\bf B}_\tau(k_\parallel,k_z)|}

\def \dE{\Sigma_\tau(k_\parallel,k_z)}

\def\tildeE{|\tilde{{\bf B}}_\tau(k_\parallel,k_z)|}

\def \kpar{k_{\parallel}}

\begin{document}

\title{Optical absorption in interacting and nonlinear Weyl semimetals}

\author{Simon Bertrand}
\author{Ion Garate}
\author{Ren\'e C\^ot\'e}
\affiliation{Institut Quantique,  Regroupement Qu\'eb\'ecois sur les Mat\'eriaux de Pointe, D\'{e}partement de Physique, Universit\'{e} de Sherbrooke, \
Sherbrooke, Qu\'{e}bec, Canada J1K 2R1}

\date{\today}

\begin{abstract}
It has been recently predicted that the interplay between Coulomb interactions and Berry curvature can produce interesting optical phenomena in topologically nontrivial two-dimensional insulators.
Here, we present a theory of the optical absorption for three-dimensional, hole-doped Weyl semimetals.
We find that the Berry curvature, Coulomb interactions and the nonlinearity in the single-particle energy spectrum can together enable a light-induced valley polarization.
We support and supplement our numerical results with an analytical toy model calculation, which unveils topologically nontrivial Mahan excitons with nonzero vorticity.
\end{abstract}

\maketitle










\section{Introduction}

The discovery of Weyl semimetals\cite{reviews} (WSM) has ignited a race of experiments aimed at identifying unambiguous physical signatures of Weyl fermions in condensed matter.
Thus far, the main efforts have been deployed towards the measurement of the chiral anomaly in electric and thermoelectric transport.\cite{huang2015, yang2015, li2016, li2016b, zhang2016, hirschberger2016} 
However, the various subleties\cite{dosreis2016, arnold2016} afflicting these experiments have put in evidence the need to develop alternative probes of Weyl fermions.  

One promising alternative route consists of measuring optical properties of WSM.
Indeed, recent theories have predicted numerous optical phenomena that originate from the hallmark energy dispersion and chirality of Weyl fermions.
To name but a few, predictions include the violation of the Wiedemann-Franz law,~\cite{tabert2016} a photoinduced anomalous Hall effect,~\cite{chan2016, song2016} a Berry-phase-induced photovoltaic effect,\cite{ishizuka2016} a quantized circular photogalvanic effect,\cite{dejuan2016} a magnetic-field-induced infared absorption from phonons,\cite{song2016b, rinkel2016}  and a magnetic-field-induced second harmonic generation.\cite{zyuzin2017} 
As of now, these predictions await experimental confirmation in spite of recent reports on related optical effects.~\cite{wu2016}

A common element to all aforementioned theoretical investigations of optical properties in WSM is that they either approximate the single-particle energy dispersion around the Weyl nodes to be perfectly linear, or they neglect electron-electron interactions. 
Hence it is natural to ask whether the interplay of band curvature, Coulomb interactions and the Berry curvature could bring about new optical effects. 
Answering this question affirmatively is the main purpose of the present work.

\begin{figure}[t]
\includegraphics[width=0.4\textwidth]{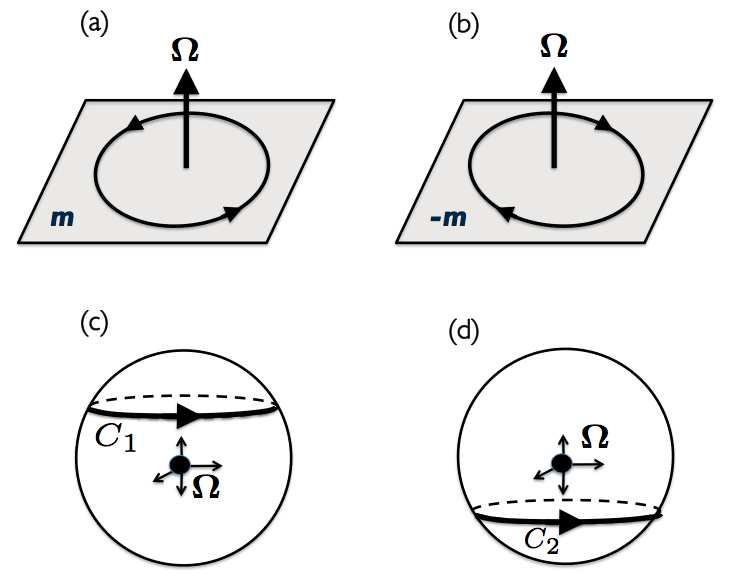}
\caption{(a) and (b): Momentum-space exciton orbits in topologically nontrivial 2D insulators. Panel (a) corresponds to an exciton state with angular momentum $m \hbar$ around the axis perpendicular to the insulator. Pannel (b) illustrates an exciton state with angular momentum $-m \hbar$. 
The Berry curvature, denoted by ${\boldsymbol\Omega}$, is perpendicular to the 2D plane and approximately constant through the exciton orbit. 
The flux of ${\boldsymbol\Omega}$ through the exciton orbit has opposite signs for $m$ and $-m$ excitons (the direction of the unit vector normal to the plane is determined by the direction of the orbit).
This difference is responsible for the chirality in the exciton spectrum.
(c) and (d): Momentum-space exciton orbits in 3D Weyl semimetals with perfectly linear energy dispersion.
In this case, the sphere denotes a constant-energy surface in momentum-space. 
The orbits $C_1$ and $C_2$ have the same orbital angular momentum around the axis that passes from the poles of the sphere.
However, the flux of the Berry curvature through both orbits has opposite signs.
In consequence, the Berry flux through an exciton orbit of given $m$ averages out to zero.
This is the reason for the absence of exciton chirality in perfectly linear Weyl semimetals.
In a Weyl semimetal with nonlinear dispersion, the constant-energy surface in momentum space is no longer spherical, and therefore the net flux through exciton orbits of a given $m$ need not average to zero.}
\label{fig:cartoon0}
\end{figure}

Our work is partly motivated by the recent literature\cite{garate2011, efimkin2013, zhou2015, srivastava2015} on the impact of the Berry curvature on excitons of two-dimensional (2D) insulators.
In topologically trivial 2D insulators, the exciton binding energy is independent of the sign of the angular momentum of the electron-hole pair about the direction perpendicular to the 2D plane. 
In contrast, in topologically nontrivial 2D insulators, the flux of the Berry curvature through the area occupied by the exciton in momentum space has opposite signs for exciton states of opposite angular momenta (see Figs.~\ref{fig:cartoon0}a and b).  
This results in a splitting of the degeneracy in their binding energies, which in turn manifests itself in a difference in the optical absorption between right- and left-circularly polarized lights (hereafter referred to as RCP and LCP, respectively).
Such phenomenon has been predicted to occur for magnetized surfaces of three dimensional topological insulators,\cite{garate2011, efimkin2013} and for $M X_2$ materials\cite{zhou2015, srivastava2015} ($M$=W, Mo and $X$=S, Se).
In this work, we wish to explore a generalization of these ideas to three dimensional (3D) WSM.

At first glance, the intended generalization is not obvious. 
In the 2D insulator, the presence of a gap in the energy spectrum  plays an essential role, for two reasons.
First, the gap is necessary in order to have a nonzero Berry curvature.
Second, the gap protects the exciton states from hybridization with the particle-hole continuum, and localizes the momentum-space wave function of the exciton in the neighborhood of the gap minimum.

Unlike the 2D insulator, a WSM has a gapless energy spectrum (barring excitonic, charge-density-wave, or related instabilities, for which no experimental evidence exists to date).
Moreover, although a  hole-doped WSM does contain an {\em optical} gap at the Fermi surface, in this case excitons are not separated from the particle-hole continuum and become resonances. 
However, these differences with respect to the 2D insulating case do not pose a serious problem, because it is sensible to calculate the effect of Coulomb interactions and Berry curvature in the optical absorption even when excitons are hybridized with the particle-hole continuum.
A more serious difference is that, unlike in 2D insulators, the Berry curvature in a WSM has the texture of a hedgehog and, accordingly, 
the flux of the Berry curvature through exciton orbits of a given angular momentum in 3D momentum space {\em changes sign} between opposite hemispheres (see Figs. ~\ref{fig:cartoon0}c and d). 
This then creates the concern that Berry curvature effects will tend to cancel out from the optical absorption, because the latter involves a sum of interband transitions over a constant energy surface in momentum space.

As it turns out, the aforementioned concern is materialized when the dispersion of the Weyl nodes is perfectly linear.
In such situation, the absorption spectra for LCP and RCP lights become identical, as if the Berry curvature effects were averaged out.
However, when (inevitable) nonlinear terms in the electronic dispersion are accounted for, the Berry curvature effect is no longer averaged out and the LCP and RCP absorption spectra become unequal.
This is the main result of our work.

The rest of this paper is organized as follows.
In Section \ref{sec:model}A, we present a model Hamiltonian for a 2-band WSM, including Coulomb interactions and the coupling to an external electric field.
As a consequence of the low-energy approximation adopted therein, the model comes with an ultraviolet energy cutoff, which is chosen to be large compared to the Fermi energy (measured from the Weyl node), but small compared to the internodal distance.
In addition, we limit ourselves to the long-ranged part of Coulomb interactions, thereby neglecting the Coulomb-interaction-induced internode scattering.
One technical advantage of this approximation is that the optical absorption of each node may be studied separately. 
This is a good approximation insofar as the nodes are sufficiently far from each other in momentum space, a circumstance that may require e.g. strong spin-orbit interactions.

In Sec. \ref{sec:model}B,  we review the formalism of the optical absorption, and apply it to a generic two-band semiconductor. 
We put particular emphasis in the discussion of the effective electron-hole interaction matrix element, which inherits information about the Berry curvature.
In Section \ref{sec:model}C, we apply the formalism of the preceding section to the nonlinear WSM introduced in Ref. [\onlinecite{ishizuka2016}].
In this nonlinear model, the Fermi surface is no longer spherically symmetric, though it maintains a cylindrical symmetry about the axis separating two neighboring Weyl nodes of opposite chirality.
In addition, we extend the model to more realistic WSM containing multiple Weyl nodes, with a focus on a time-reversal-symmetric WSM and an inversion-symmetric WSM.
In both cases, we assume the presence of at least a mirror plane, which is a common occurrence in most WSM.
One important result of this section is that the optical absorption for RCP light
involves particle-hole pairs with angular momentum $1$ (in units of $\hbar$) around the axis of the cylinder, while the optical absorption for LCP light involves electron-hole pairs of angular momentum $-1$.
These selection rules, which hold so long as the propagation direction of the light is parallel to the axis of cylindrical symmetry,  are central to the main results of this paper.

Section \ref{sec:numerical} is devoted to numerical results.
The first main finding is that LCP and RCP absorption spectra are degenerate in the perfectly linear WSM model.
We attribute such degeneracy to a pseudo time-reversal symmetry that emerges in the linear spectrum approximation.
The nonlinear terms in the spectrum break this symmetry, and consequently LCP and RCP absorption spectra become non degenerate.
Roughly speaking, the nonlinearity in the single-particle spectrum enables the Berry curvature to manifest itself in the optical absorption spectrum.
The difference between the LCP and RCP absorption spectra (which we variously refer to as the RCP-LCP splitting/asymmetry/difference) scales with the frequency of the absorbed photon. 
This is a direct consequence of the fact that higher-frequency photons excite electron-hole pairs of higher momenta (where band curvature effects are more pronounced).
In a WSM with multiple Weyl nodes, the combination of Coulomb interactions, nonlinearity and Berry phase results in a light-induced valley polarization.
Although valley polarization has been amply studied in graphene\cite{rycerz2007} and topologically nontrivial 2D insulators,\cite{dai2012} to the best of our knowledge the prediction of valley polarization in WSM is new.

Finally, Sec.~\ref{sec:analytical} is devoted to an approximate analytical solution of the problem, based on the replacement of the Coulomb potential by a contact interaction.
The aim of this section is to corroborate and better understand the numerical results of the preceding section.
Simple analysis shows that the electron-hole pairs near the absorption threshold can be regarded as topological Mahan excitons:\cite{mahan1966} they have exponentially small binding energies and contain nodes with nonzero vorticity.
Moreover, the analytical solution allows to relate the asymmetry between the LCP and RCP absorption spectra to the Berry curvature.
Specifically, the asymmetry emerges from a nonzero average over the Fermi surface of the component of the Berry curvature along the direction separating two neighboring nodes with opposite chirality.
Nonlinear terms in the energy dispersion are essential in order to have a nonzero value for said average.

Concerning notation, we take $\hbar\equiv 1$ and SI units throughout.

\section{Model and formalism}
\label{sec:model}

\subsection{Hamiltonian}

\begin{figure}[t]
\includegraphics[width=0.2\textwidth]{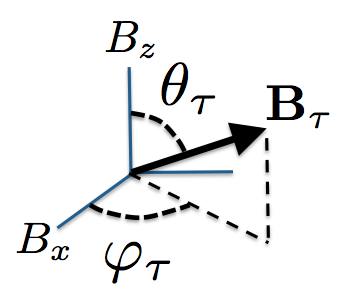}
\caption{The effective magnetic field acting on the pseudospin space at node $\tau$.}
\label{fig:Bfig}
\end{figure}

In the low-energy approximation, a WSM is characterized by a set of Weyl nodes, which we label with an index $\tau$. 
In the absence of Coulomb interactions, the electronic structure around a node $\tau$ is described by an effective two-band ${\bf k}\cdot{\bf p}$ Hamiltonian
\begin{equation}
\label{eq:h}
h_\tau({\bf k})=C_\tau({\bf k})+{\bf B}_\tau({\bf k})\cdot{\boldsymbol\sigma},
\end{equation}
where ${\boldsymbol\sigma}$ is a pseudospin denoting the two bands that touch at the Weyl node, ${\bf k}$ is the wave vector measured from the node, and ${\bf B}_\tau({\bf k})$ is an effective magnetic field acting on the pseudospin space.
This model is valid for $|{\bf B}_\tau({\bf k})|<\Lambda$, where $\Lambda$ is an ultraviolet energy cutoff such that ${\bf k}$ is small compared to the internodal distance.
The eigenvectors of $h_\tau({\bf k})$ are
\begin{align}
\label{eq:ckvk}
| {\bf k} c \tau\rangle &=  \left(\begin{array}{c} \cos\frac{\theta_{{\bf k}\tau}}{2} \\
                                                                         e^{i\varphi_{{\bf k}\tau}} \sin\frac{\theta_{{\bf k}\tau}}{2}
\end{array}\right)\nonumber\\
| {\bf k} v \tau\rangle &=  \left(\begin{array}{c} -\sin\frac{\theta_{{\bf k}\tau}}{2} \\
                                                                         e^{i\varphi_{{\bf k}\tau}} \cos\frac{\theta_{{\bf k}\tau}}{2}
\end{array}\right),
\end{align}
where $c$ and $v$ stand for the conduction and valence band, respectively, whereas $\theta_{{\bf k}\tau}$ and $\varphi_{{\bf k}\tau}$ are the polar and azimuthal angles of the vector ${\bf B}_\tau({\bf k})$ (see Fig.~\ref{fig:Bfig}).
The corresponding eigenvalues are $E_{{\bf k}c\tau} = C_\tau({\bf k})+|{\bf B}_\tau({\bf k})|$ and $E_{{\bf k} v \tau} = C_\tau({\bf k}) -|{\bf B}_\tau({\bf k})|$.
In the second quantized form, the noninteracting model can thus be written as
\begin{equation}
{\cal H}_0 = \sum_{{\bf k} \alpha \tau} E_{{\bf k}\alpha\tau} c^\dagger_{{\bf k}\alpha\tau} c_{{\bf k}\alpha\tau},
\end{equation} 
where $\alpha=c, v$ and $c^\dagger_{{\bf k}\alpha\tau}$ is an operator that creates an electron in state $|{\bf k}\alpha\tau\rangle$.

In this work, we wish to investigate the influence of electron-electron interactions in the optical absorption.
In the second quantized form, the Coulomb interaction reads
\begin{equation}
\label{eq:U}
{\cal U}=\frac{1}{2}\int d^3 r d^3 r' V_{\rm sc}({\bf r}-{\bf r}')\Psi^\dagger({\bf r}) \Psi^\dagger({\bf r}')\Psi({\bf r}') \Psi({\bf r}),
\end{equation}
where 
$V_{\rm sc}({\bf r})$
is the screened Coulomb potential. 
The field operators in Eq.~(\ref{eq:U}) can be expanded onto the band eigenstates near the Weyl nodes,
\begin{equation}
\label{eq:Psi}
\Psi({\bf r}) \simeq \frac{1}{\sqrt{\cal V}}\sum_{{\bf k}\lambda\tau} e^{i {\bf k}_\tau\cdot{\bf r}} e^{i{\bf k}\cdot{\bf r}} |{\bf k}\lambda\tau\rangle c_{{\bf k}\lambda\tau},
\end{equation}
where $\lambda\in\{c, v\}$, ${\cal V}$ is the volume of the sample, ${\bf k}_\tau$ is the location of node $\tau$ in momentum space and $ {\bf k}$ is the wave vector measured from the node.
Substituting Eq.~(\ref{eq:Psi}) in Eq.~(\ref{eq:U}), we get
\begin{align}
\label{eq:U2}
{\cal U}\simeq&\frac{1}{2 {\cal V}}\sum_{\lambda \lambda' \gamma \gamma'}\sum_{{\bf k}{\bf k}'{\bf q}} \sum_{\tau\tau'} \langle {\bf k}+{\bf q} \lambda \tau|{\bf k} \gamma \tau\rangle\langle{\bf k}'-{\bf q}\lambda' \tau' |{\bf k}' \gamma' \tau'\rangle 
\nonumber\\
&~~~~~~~~~~~\times V_{\rm sc}({\bf q})c^\dagger_{{\bf k}+{\bf q} \lambda\tau} c^\dagger_{{\bf k}'-{\bf q}\lambda'\tau'} c_{{\bf k}'\gamma'\tau'} c_{{\bf k}\gamma \tau},
\end{align}
where 
\begin{equation}
V_{\rm sc}({\bf q})= \frac{e^2}{\epsilon_0 \epsilon_\infty q^2 \epsilon({\bf q})}
\end{equation}
is the Fourier transform of $V_{\rm sc}(r)$, $\epsilon_0$ is the vacuum permittivity, $\epsilon_\infty$ is the contribution to the dielectric constant coming from the high-energy bands not included in Eq.~(\ref{eq:h}), and $\epsilon({\bf q})$ is the static dielectric function originating from particle-hole excitations in the two-band model.
In the derivation of Eq.~(\ref{eq:U2}), we have neglected the Fourier modes of $V_{\rm sc}({\bf q})$ involving values of $q$ larger than the high-energy cutoff.
This approximation is motivated by the fact that the optical absorption of weakly doped WSM is dominated by the long-wavelength part of the Coulomb interaction.  
Consequently, internode scattering produced by Coulomb interactions is neglected and all momenta appearing in Eq.~(\ref{eq:U2}) have cutoffs.

\begin{figure}[t]
\includegraphics[width=0.4\textwidth]{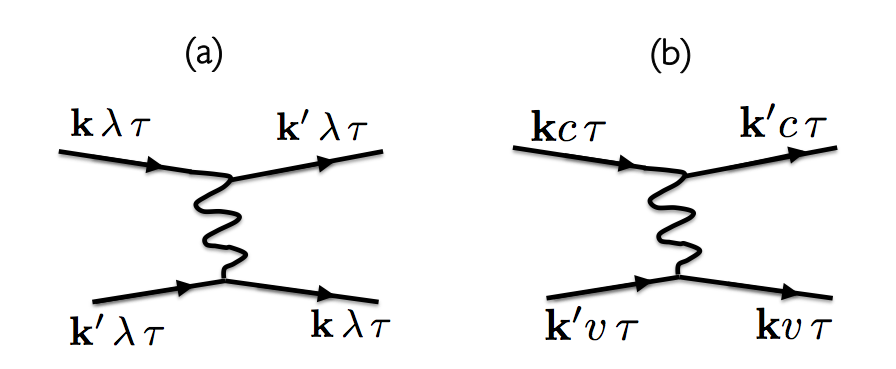}
\caption{Feynman diagrams for the Coulomb interaction terms considered in the main text. The index $\lambda$ stand for the conduction or valence band. The diagram (a) renormalizes the optical gap, while the diagram (b) leads to electron-hole pairing.}
\label{fig:coulomb}
\end{figure}

There is one more approximation to be made for ${\cal U}$.
Namely, we are to neglect interband Coulomb scattering (from the conduction to the valence band or vice versa), which is justified based on the facts that 
(i) the Coulomb interaction is maximal at small momentum transfer between the scattered electrons, (ii) the overlap between Bloch spinors at the same momenta and different bands vanishes.
This then leaves us with the Coulomb scattering processes depicted in Fig.~\ref{fig:coulomb}.
Similar approximations are common in textbook discussions of the optical absorption.\cite{haug2009}

The last ingredient of the model is the coupling between electrons and the electromagnetic field.
In the ``length gauge'',\cite{aversa1995} we have
\begin{equation}
\label{eq:he0}
{\cal H}_{\cal E} = \int d^3 r \Psi^\dagger({\bf r}) (-e {\bf r})\cdot\boldsymbol{{\cal E}}(t) \Psi({\bf r}),
\end{equation}
where $\boldsymbol {{\cal E}}(t)$ is the electric field (approximately uniform) corresponding to the incident light.
Adopting the low-energy and dipole approximations, and keeping only interband terms (which are the ones that participate in optical absorption),  Eq.~(\ref{eq:he0}) becomes
\begin{equation}
\label{eq:he}
{\cal H}_{\cal E} = {\cal E}(t) \sum_{{\bf k}\tau} {\bf d}_\tau({\bf k}) c^\dagger_{{\bf k} c\tau} c_{{\bf k} v \tau} + {\rm h. c},
\end{equation}
where ${\bf k}$ is measured with respect to the nodes,
\begin{equation}
{\bf d}_\tau({\bf k}) = i e \langle {\bf k} c \tau | \grad_{\bf k} | {\bf k} v \tau\rangle = \frac{i e  \langle {\bf k} c \tau | {\bf v} | {\bf k} v \tau\rangle}{E_{{\bf k} v \tau}-E_{{\bf k} c \tau}} 
\end{equation}
is the interband dipole matrix element and ${\bf v}=\partial {\cal H}_0/\partial {\bf k}$ is the velocity operator of noninteracting electrons.
The full Hamiltonian that we will consider is thus
\begin{equation}
{\cal H}={\cal H}_0+{\cal U} +{\cal H}_{\cal E}(t).
\end{equation}

\subsection{Optical absorption}

\begin{figure}[t]
\includegraphics[width=0.4\textwidth]{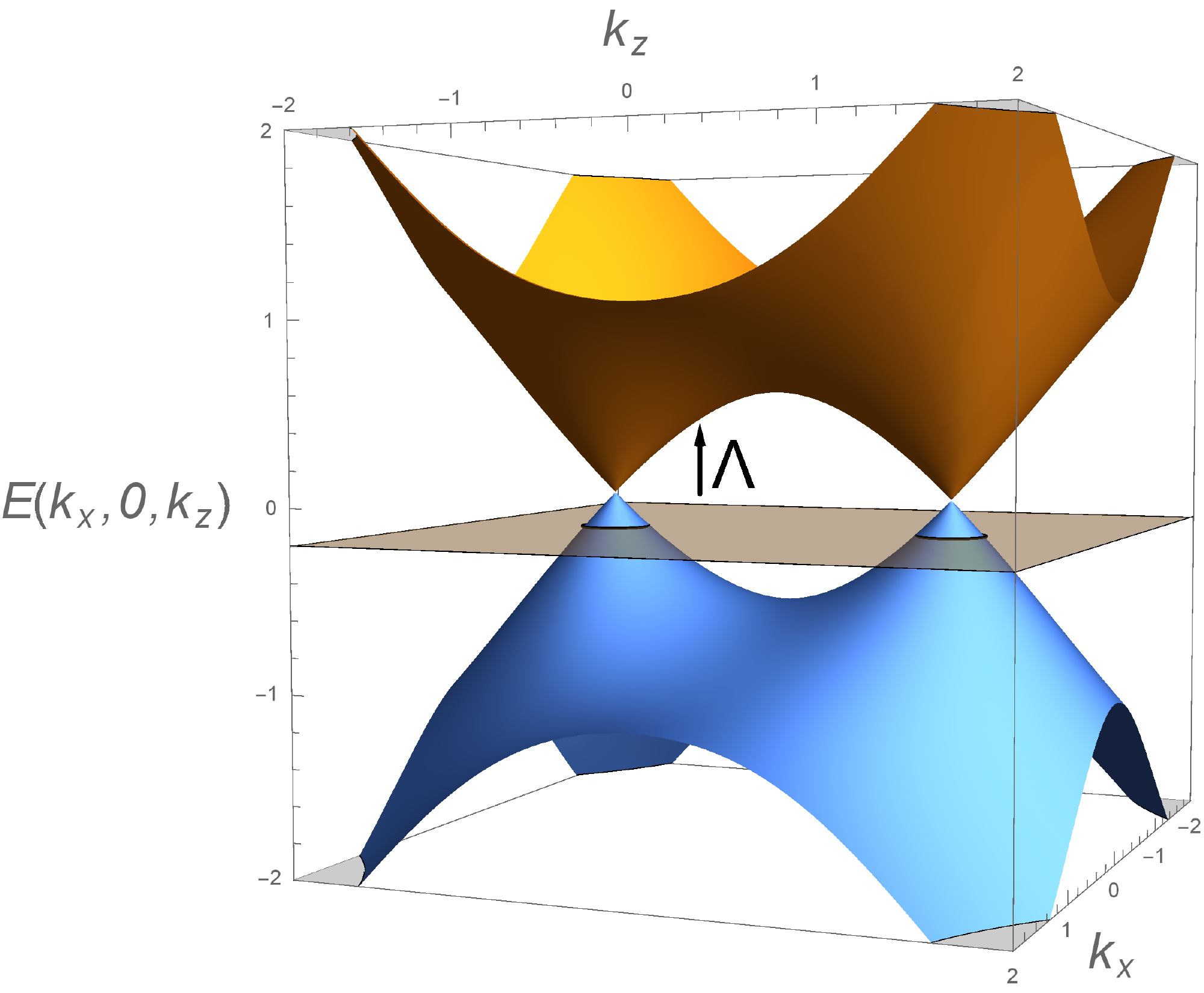}
\caption{Energy dispersion in the vicinity of two Weyl nodes of opposite chirality. 
 We consider electronic states within an energy interval $(-\Lambda, \Lambda)$.
The Fermi energy (the constant energy plane in the figure) is located within the interval and intersects with the valence band.}
\label{fig:disp}
\end{figure}

The main objective of this work is to investigate the optical absorption of a hole-doped WSM (see Fig.~\ref{fig:disp}).
The central quantity in the optical absorption is the macroscopic interband polarization ${\bf P}(t)$ (dipole moment per unit volume) defined as 
\begin{equation}
{\bf P}(t) = \frac{1}{\cal V}\sum_{{\bf k} \tau} P_\tau({\bf k}, t) {\bf d}_\tau({\bf k}) +{\rm c.c},
\end{equation}
where 
\begin{equation}
\label{eq:coh}
P_\tau({\bf k}, t) = \langle c^\dagger_{{\bf k} v \tau}(t) c_{{\bf k} c \tau }(t)\rangle
\end{equation}
is the (dimensionless) interband coherence, $c_{{\bf k} \alpha\tau}(t) = \exp(i{\cal H} t) c_{{\bf k}\alpha\tau} \exp(-i{\cal H} t)$ and the average in Eq.~(\ref{eq:coh}) is taken over the ground state of ${\cal H}_0+{\cal U}$.
In equilibrium and in absence of an excitonic condensate, $P_\tau({\bf k}, t)=0$.
However, under light irradiation,  $P_\tau({\bf k}, t)\neq 0$, which then determines the optical absorption coefficient.
In order to calculate $P_\tau({\bf k}, t)$, we follow the equation of motion approach from Ref.~[\onlinecite{haug2009}] and arrive at
\begin{widetext}
\begin{equation}
\label{eq:P2b}
\left [(\omega+ i\delta) -2 |{\bf B}_\tau({\bf k})|  - \Sigma_{{\bf k}\tau}\right] P_\tau({\bf k},\omega) = (f_{c \tau}({\bf k})- f_{v\tau}({\bf k}))\left( \boldsymbol{{\cal E}}(\omega)\cdot {\bf d}_\tau({\bf k}) +\frac{1}{\cal V}\sum_{\bf k'} V_\tau({\bf k}, {\bf k}')P_\tau ({\bf k}',\omega)\right),
\end{equation}
where $\omega$ is the frequency of the electric field, $f_{v \tau}({\bf k})$ and $f_{c \tau}({\bf k})$ are the single-particle occupation factors, $\delta$ is an adiabatic switch-on factor to ensure that ${\cal E}(t)\to 0$ when $t\to-\infty$,
\begin{equation}
V_\tau({\bf k},{\bf k}') = V_{\rm sc}({\bf k}-{\bf k}) \langle {\bf k} c\tau | {\bf k}' c \tau\rangle \langle {\bf k}' v \tau | v {\bf k} \tau\rangle
\end{equation}
is the Coulomb interaction including the band eigenstate overlap matrix elements, and
\begin{align}
\Sigma_\tau({\bf k}) &= \frac{1}{\cal V}\sum_{\bf k'} V_{\rm sc}({\bf k}-{\bf k}')  |\langle {\bf k} v \tau | {\bf k}' v \tau\rangle|^2 f_{v\tau}({\bf k}')
\end{align}
is the {\em difference} between the conduction and valence band self-energies, which renormalizes the optical gap.
For brevity, we will refer to it as the self-energy.
In the derivation of Eq.~(\ref{eq:P2b}), we have assumed that there is no internode coherence induced by the light ($\langle c^\dagger_{{\bf k} v \tau} c_{{\bf k} c\tau'}\rangle = 0$ for $\tau\neq \tau'$).
Accordingly, the optical absorption of each node can be studied separately.
In addition, we have adopted the quasi-equilibrium approximation,\cite{haug2009} so that $f_{\alpha\tau}({\bf k})$ are time-independent Fermi-Dirac distributions with an effective Fermi energy.
In the linear response approximation pursued below, these occupation factors will be taken equal to those in absence of light.
Note that $C_\tau({\bf k})$ is implicitly present in Eq.~(\ref{eq:P2b}) through $f_{\alpha\tau}({\bf k})$.

The quantity $\langle c{\bf k}\tau|c {\bf k}'\tau\rangle\langle v{\bf k}'\tau|v {\bf k}\tau\rangle$ appearing in the Coulomb interaction is not gauge-invariant, though, of course, all physical observables (like the optical absorption) will be independent of the gauge choice.
Our gauge choice is set by Eq.~(\ref{eq:ckvk}), which yields
\begin{equation}
\label{eq:me}
\langle c{\bf k}\tau|c {\bf k}'\tau\rangle\langle v{\bf k}'\tau|v {\bf k}\tau\rangle = \frac{1}{2}\left[\sin\theta_\tau\sin\theta'_\tau+(1+\cos\theta_\tau\cos\theta'_\tau)\cos(\varphi_\tau-\varphi'_\tau)+i(\cos\theta_\tau+\cos\theta_\tau)\sin(\varphi_\tau-\varphi'_\tau)\right].
\end{equation}
\end{widetext}
For brevity, we denote $\theta_{{\bf k}\tau} (\varphi_{{\bf k}\tau})$ and $\theta_{{\bf k}'\tau} (\varphi_{{\bf k}'\tau})$ as $\theta_\tau (\varphi_\tau)$ and $\theta'_\tau (\varphi'_\tau)$, respectively.

\begin{figure}[t]
\includegraphics[width=0.2\textwidth]{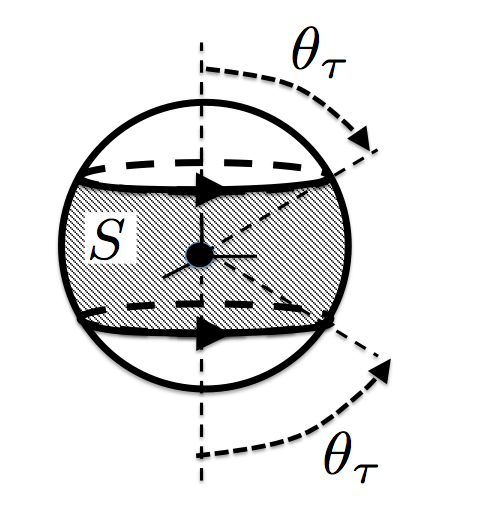}
\caption{A sphere in ${\bf B}$-space, centered at the Weyl node $\tau$. 
The factor $\cos\theta_\tau$ referred to in the main text can be interpreted geometrically as the flux of the joint Berry curvature defined in Eq.~(\ref{eq:berrycurv}) through the shaded surface $S$:
$\cos\theta_\tau = (1/4\pi)\int_S {\boldsymbol\Omega}_{\rm Sch}\cdot d{\bf S}$, where $d{\bf S}= |{\bf B}_\tau|^2 \sin\theta_\tau d\theta_\tau d\varphi\, \hat{\bf B}$.
The flux is positive (negative) when $\theta_\tau<\pi/2$ ($\theta_\tau>\pi/2$).
It will be shown in Sec.~\ref{sec:analytical} that the average of $\cos\theta_\tau$ over a constant-energy surface in momentum space determines the difference between the absorption spectra of left- and right-circularly polarized lights. 
In a Weyl semimetal with perfectly linear spectrum, the average vanishes. In contrast, the average is rendered nonzero by the presence of nonlinear terms in the spectrum.}
\label{fig:sphere}
\end{figure}

Since the Coulomb interaction is strongest when ${\bf k}\simeq {\bf k}'$, we analyze the Coulomb matrix elements in that regime.
We get
\begin{align}
\langle c {\bf k} \tau | c {\bf k}' \tau\rangle & \simeq \exp \left(i\sin^2\frac{\theta_\tau}{2} \delta\varphi_\tau\right)=\exp\left(-i {\bf A}_c\cdot \delta{\bf B}_\tau\right)\nonumber\\
\langle v {\bf k}' \tau | v {\bf k} \tau\rangle & \simeq \exp \left(-i\cos^2\frac{\theta_\tau}{2} \delta\varphi_\tau\right)=\exp\left(i {\bf A}_v\cdot \delta {\bf B}_\tau\right),
\end{align}
where $\delta\varphi_\tau=\varphi'_\tau-\varphi_\tau$, $\delta {\bf B}_\tau= {\bf B}_\tau({\bf k}') - {\bf B}_\tau({\bf k})$ and
${\bf A}_{c (v)} = i \langle {\bf k} c(v) \tau| \grad_{\bf B}|{\bf k} c(v) \tau\rangle$ are the Berry connections for the conduction and valence bands.
Note that these connections are defined with respect to ${\bf B}$ rather than ${\bf k}$.
Explicitly, 
\begin{align}
{\bf A}_c &= \frac{-1+\cos\theta_\tau}{2}\grad_{\bf B}\varphi_\tau = \frac{-1+\cos\theta_\tau}{2 |{\bf B}_\tau| \sin\theta_\tau} \hat{\boldsymbol{\varphi}}\nonumber\\
{\bf A}_v &= \frac{-1-\cos\theta_\tau}{2}\grad_{\bf B}\varphi_\tau = \frac{-1-\cos\theta_\tau}{2 |{\bf B}_\tau|\sin\theta_\tau} \hat{\boldsymbol{\varphi}}
\end{align}
represent the gauge fields created by a monopole located at the Weyl node.
Then, Eq.~(\ref{eq:me}) becomes
\begin{equation}
\langle c{\bf k}\tau|c {\bf k}'\tau\rangle\langle v{\bf k}'\tau|v {\bf k}\tau\rangle \simeq e^{-i \cos\theta_\tau \delta\varphi_\tau} \,\,\text{(for ${\bf k}\simeq {\bf k}'$}).
\end{equation}
This is nothing but the phase of a particle moving on a Schwinger vector potential\cite{shnir2005} 
\begin{equation}
\label{eq:schwinger}
{\bf A}_{\rm Sch} \equiv {\bf A}_c-{\bf A}_v=\frac{1}{|{\bf B}_\tau|} \frac{\cos\theta_\tau}{\sin\theta_\tau} \hat{\boldsymbol{\varphi}},
\end{equation}
which in our case is the ``joint'' (particle-hole) Berry connection.
The factor $\cos\theta_\tau$, which will play an important role in our results, can be associated with the flux of the ``joint'' Berry curvature 
\begin{equation}
\label{eq:berrycurv}
\boldsymbol{\Omega}_{\rm Sch} = \grad_{{\bf B}_\tau} \times {\bf A}_{\rm Sch} =- {\bf B}_\tau/B_\tau^3 
\end{equation}
through the surface shown in Figure~\ref{fig:sphere}.

One of the main drives of our work is to evaluate how the Berry phase appearing in the Coulomb matrix elements impacts the optical absorption of a WSM.
In topologically nontrivial two dimensional systems, such as the surface of a magnetized topological insulator\cite{garate2011} or $M X_2$ compounds\cite{zhou2015, srivastava2015} ($M$=Mo, W; $X$=S, Se), a true energy gap in the spectrum is essential in order to have a nonzero Berry curvature.
Furthermore, the gap ensures that the exciton wave function is peaked near the bandgap minimum, where  $\cos\theta\simeq \pm 1$ (the sign depends on the sign of the Berry curvature).
In such systems, the effect of the Schwinger potential amounts to shifting\cite{efimkin2013} the azimuthal angular momentum of the electron-hole pair by $\hbar$, i.e., $\exp(i m\varphi) \to \exp[i (m\pm 1)\varphi)]$, thereby leading to chiral excitons.\cite{garate2011} 
In our case, there is no true energy gap in the spectrum of the WSM, but instead we have an optical gap at the Fermi surface.
Moreover, the value of $\cos\theta_\tau$ at the Fermi surface can take both positive and negative values.
Consequently, the $\cos\theta_\tau$ factor tends to average out from the theory and one may expect that the effect of the Berry phase in Coulomb matrix elements will not impact the optical absorption of a WSM in a qualitative manner.
Nevertheless, as we shall show below, this expectation holds only for a WSM with a perfectly linear energy spectrum.
The inevitable nonlinearities in the energy spectrum will prevent the complete averaging-out of the Schwinger potential, and will translate into an asymmetry between the optical absorption spectra of LCP and RCP lights.

The standard\cite{haug2009} strategy to solve Eq.~(\ref{eq:P2b}) is to first to expand $P_\tau({\bf k},\omega)$ onto an orthonormal basis as 
\begin{equation}
\label{eq:expansion}
P_\tau({\bf k}, \omega) = \sum_n a_{n\tau} (\omega) \psi_{n\tau} ({\bf k}),
\end{equation} 
where $a_{n\tau}$ are complex coefficients to be determined and the function $\psi_{n\tau}({\bf k})$ satisfies a Wannier equation
\begin{align}
\label{eq:wannier}
&\left(\epsilon_{n\tau}-2 |{\bf B}_\tau({\bf k})| - \Sigma_\tau({\bf k})\right) \psi_{n\tau}({\bf k})\nonumber\\
& = (f_{c \tau}({\bf k})- f_{v\tau}({\bf k})) \frac{1}{\cal V}\sum_{\bf k'} V_\tau({\bf k}, {\bf k}')\psi_{n\tau}({\bf k}').
\end{align}
This equation can be interpreted as an effective Schr\"odinger equation for a particle-hole pair with excitation energies $\epsilon_{n\tau}$ and wave functions $\psi_{n\tau}({\bf k})$, where $n$ is the eigenvalue index.
A similar equation may be derived from the Green's function approach.\cite{onida2002}
The numerical and (approximate) analytical solutions of Eq.~(\ref{eq:P2b}) will be discussed in Secs.~\ref{sec:numerical} and \ref{sec:analytical}, respectively.
For now, let us suppose that the eigenvalues $\epsilon_{n\tau}$ and the eigenfunctions $\psi_{n\tau}({\bf k})$ are known.
Combining Eqs.~(\ref{eq:P2b}), (\ref{eq:expansion}) and (\ref{eq:wannier}),  and using $\sum_{\bf k} \psi^*_{n\tau}({\bf k}) \psi_{n'\tau} ({\bf k}) = \delta_{n n'}$, we obtain
\begin{align}
\label{eq:Pk}
&P_\tau({\bf k},\omega)=\sum_n \frac{\psi_{n\tau}({\bf k})}{(\omega+i\delta) - \epsilon_{n\tau}}\nonumber\\
&\times \sum_{\bf k'} (f_{c\tau}({\bf k}')-f_{v\tau}({\bf k}'))  \psi^*_{n\tau}({\bf k}') {\bf d}_\tau ({\bf k}')\cdot\boldsymbol{{\cal E}}(\omega).
\end{align}
The valley-resolved interband polarization can now be written as 
\begin{equation}
\label{eq:pom}
{\bf P}_\tau(\omega) =\frac{1}{\cal V}\sum_{\bf k} \left[P_\tau({\bf k},\omega) {\bf d}^*_\tau({\bf k}) + P^*_\tau({\bf k},-\omega) {\bf d}_\tau({\bf k})\right],
\end{equation}
while the full interband polarization reads ${\bf P}(\omega)=\sum_\tau {\bf P}_\tau(\omega)$.
In linear response theory, it is customary to rewrite Eq.~(\ref{eq:pom}) as
\begin{equation}
{\bf P}_\tau(\omega) = \epsilon_0 {\boldsymbol \chi}_\tau(\omega) \cdot \boldsymbol{{\cal E}}(\omega),
\end{equation}
where ${\boldsymbol \chi}_\tau(\omega)$ is the (dimensionless) valley-resolved electric susceptibility tensor.
Below, we will be interested in the absorptive (imaginary) part of the susceptibility, $\boldsymbol{\chi}_\tau''$.
The eigenvalues of $\boldsymbol{\chi}_\tau''$, denoted as $\chi_{\tau l}''$, give the optical absorption coefficients (in units of inverse length) for node $\tau$:
\begin{equation}
\alpha_{\tau l}(\omega)=\frac{\omega}{n' c} \chi_{l\tau}''(\omega),
\end{equation}
where $l=1,2,3$ is the eigenvalue index, $c$ is the speed of light and $n'$ is the refractive index of the material (whose frequency-dependence may be neglected in the range of interest).
The total absorption coefficient $\alpha_l(\omega)=\sum_\tau \alpha_{l\tau}(\omega)$ can be determined experimentally via reflectivity measurements.\cite{cardona2010}
The calculation of valley-resolved absorption coefficients will be the main objective of Secs.~\ref{sec:numerical} and \ref{sec:analytical}.

\subsection{Application to nonlinear WSM}

Thus far, the formalism presented has been valid for a generic two-band model with multiple valleys.
Here, we discuss the case of a WSM in more detail.
The non-interacting Hamiltonian near one of the nodes (e.g., $\tau\equiv 1$) is characterized by the toy model\cite{ishizuka2016}
\begin{align}
\label{eq:B1}
B_{1,x}({\bf k}) &=  v k_x (1+\alpha k_z)\nonumber\\
B_{1,y}({\bf k}) &=  v k_y (1+\alpha k_z)\nonumber\\
B_{1,z}({\bf k}) &=  v_z k_z + \beta (\kpar^2 - 2 k_z^2),
\end{align}
where ${\bf k} = ({\bf \kpar},k_z)$, ${\bf\kpar}=(k_x,k_y)$ and $v$ and $v_z$ are the Dirac velocities.
The parameters $\alpha$ and $\beta$ account for the leading nonlinear corrections to the Weyl Hamiltonian (note that $\alpha$ and $\beta$ have different dimensions).
The parameter $\alpha$ is not to be confused with the optical absorption coefficient $\alpha(\omega)$; we will attach the frequency label only to the latter.
As we shall see, the nonlinear terms in the single-particle energy spectrum alter the optical properties of the WSM qualitatively.
Equation (\ref{eq:B1}) displays a cylindrical symmetry around the $k_z$ direction. 
Physically, $k_z$ is the direction that separates a pair of Weyl nodes of opposite chirality. 
In Eq.~(\ref{eq:ckvk}), $\varphi_{{\bf k}\tau}$ coincides with the azimuthal angle of the wave vector $ {\bf k}$. 
However, $\theta_{{\bf k}\tau}$ differs from the polar angle of ${\bf k}$ whenever $\alpha$ or $\beta$ are nonzero.

The nonlinear Weyl model is valid only at low energies, $|{\bf B}_\tau({\bf k})|<\Lambda$, where $\Lambda$ is an ultraviolet (UV) cutoff.
Consequently, all momenta appearing in Eq.~(\ref{eq:B1}) have UV cutoffs (which are not symmetric about $k=0$ in presence of nonlinear terms).
The energy cutoff is chosen to be large compared to $\epsilon_F$, the Fermi energy measured from the Weyl node ($\epsilon_F<0$ for a hole-doped WSM). 
In addition, the cutoff must be small enough so that the nonlinear terms in the dispersion are subdominant with respect to the linear ones.

In a WSM, Weyl nodes appear in pairs of opposite chirality.
In this work, we shall be interested in two cases: (i) WSM with time-reversal (TR) symmetry (and broken inversion symmetry), (ii) WSM with inversion (I) symmetry (and broken TR symmetry).
For both cases, we shall assume that the crystal has at least one mirror symmetry, which is a common circumstance.

In a WSM with time-reversal symmetry, there must be at least four nodes (unless a node occurs at a time-reversal-invariant momentum, a situation that we do not consider here). 
We adopt the minimal case, i.e., four nodes, though the generalization to more nodes is straightforward. 
Nodes 1 and 2 are related to one another by a mirror plane,
\begin{equation}
h_1({\cal M}{\bf k}) = {\cal M}^{-1} h_2({\bf k}) {\cal M},
\end{equation}
while nodes 3 and 4 are the time-reversed partners of nodes 1 and 2, respectively, e.g.
\begin{equation}
h_3({\cal T} {\bf k}) = {\cal T}^{-1} h_1({\bf k}) {\cal T} \text{  (TR-symmetric WSM)}.
\end{equation}
In a non-centrosymmetric material with spin-orbit coupling, the pseudospin $\sigma$ will transform like a spin under time reversal and mirror operations.\cite{hirayama2015}
Because the $z$ direction in the nonlinear model is the one separating a pair of nodes of opposite chirality, we take a mirror plane perpendicular to $z$: ${\cal M}=i\sigma_z\otimes (k_z\to -k_z)$.
In addition, 
${\cal T}=i \sigma_z K \otimes ({\bf k}\to -{\bf k})$, where $K$ is the complex conjugation.
Table 1 lists the form of ${\bf B}_\tau({\bf k})$ for the different nodes.
In addition, these symmetries impose $C_1({\bf k})=C_2({\cal M} {\bf k})$, $C_3({\bf k})=C_1(-{\bf k})$ and $C_4({\bf k})=C_2(-{\bf k})$.
Accordingly, $C_\tau({\bf 0})$ is the same for all $\tau$, i.e., the four Weyl nodes are at the same energy.
In addition, for the sake of concreteness, we will hereafter neglect the momentum-dependence of $C_\tau({\bf k})$.
Thus, we will neglect the tilt of Weyl nodes and our results will be focused on the simplest WSM of type I.
In practice, this means that $C_\tau$ will disappear from Eqs.~(\ref{eq:P2b}) and (\ref{eq:wannier}).

A minimal WSM with inversion symmetry has two Weyl nodes, but for consistency we consider the case of four nodes here too. 
Nodes 1 and 2 are related to one another the mirror plane ${\cal M}$, while nodes 3 and 4 are the inversion partners of nodes 1 and 2, respectively, e.g. 
\begin{equation}
h_3({\cal P} {\bf k}) = {\cal P}^{-1} h_1({\bf k}) {\cal P} \text{  (I-symmetric WSM)}.
\end{equation}
Here, ${\cal P}$ is the inversion operator, which takes ${\bf k}\to -{\bf k}$ and acts as an identity in $\sigma$ space.
Table 1 lists the form of ${\bf B}_\tau({\bf k})$ for the different nodes.
The symmetry relations for $C_\tau({\bf k})$ are identical to the ones from the preceding paragraph.
Accordingly, the four Weyl nodes are at the same energy in this case as well.

Real WSM often have different sets of Weyl nodes at different energies.
Our model captures a set of equienergetic Weyl nodes that are closest to the Fermi energy.
The sets of Weyl nodes that are further away from the Fermi energy will have their optical absorption thresholds at higher frequencies, and thus their contribution can be separated out.

\begin{table*}[tb]
\caption{Model Hamiltonians for four Weyl nodes (labelled $\tau=1,2, 3, 4$) in a WSM with time-reversal symmetry.
Nodes 1 and 2 are related to one another by a mirror plane perpendicular to $k_z$.
Nodes 3 and 4 are the time-reversed partners of nodes 1 and 2, respectively. 
}\label{tab:TRS}
\begin{ruledtabular}
\begin{tabular}{ccccc}
Nodes  & $\tau=1$  &  $\tau=2$ & $\tau=3$ & $\tau=4$ \\ \hline
$B_{\tau, x}({\bf k})$ & $ v k_x (1+\alpha k_z)$ & $- v k_x (1-\alpha k_z)$ & $ v k_x (1-\alpha k_z)$ & $- v k_x (1+\alpha k_z)$ \\ \hline
$B_{\tau, y}({\bf k})$ &  $ v k_y (1+\alpha k_z)$ &  $- v k_y (1-\alpha k_z)$ & $ v k_y (1-\alpha k_z)$ & $- v k_y (1+\alpha k_z)$ \\ \hline
$B_{\tau, z}({\bf k})$ &  $ v_z k_z + \beta (\kpar^2- 2 k_z^2)$ & $- v_z k_z + \beta (\kpar^2 - 2 k_z^2)$ & $ v_z k_z - \beta (\kpar^2 - 2 k_z^2)$ & $- v_z k_z - \beta (\kpar^2 - 2 k_z^2)$
\end{tabular}
\end{ruledtabular}
\end{table*}

\begin{table*}[tb]
\caption{Model Hamiltonians for four Weyl nodes (labelled $\tau=1,2, 3, 4$) in a WSM with inversion symmetry.
Nodes 1 and 2 are related to one another by a mirror plane perpendicular to $k_z$.
Nodes 3 and 4 are the space-inversion partners of nodes 1 and 2, respectively. 
}\label{tab:P}
\begin{ruledtabular}
\begin{tabular}{ccccc}
Nodes  & $\tau=1$   &  $\tau=2$  & $\tau=3$  & $\tau=4$ \\ \hline
$B_{\tau, x}({\bf k})$ & $ v k_x (1+\alpha k_z)$ & $- v k_x (1-\alpha k_z)$ & $- v k_x (1-\alpha k_z)$ & $ v k_x (1+\alpha k_z)$ \\ \hline
$B_{\tau, y}({\bf k})$ &  $ v k_y (1+\alpha k_z)$ &  $- v k_y (1-\alpha k_z)$ & $- v k_y (1-\alpha k_z)$ & $ v k_y (1+\alpha k_z)$ \\ \hline
$B_{\tau, z}({\bf k})$ &  $ v_z k_z + \beta (\kpar^2- 2 k_z^2)$ & $- v_z k_z + \beta (\kpar^2 - 2 k_z^2)$ & $- v_z k_z + \beta (\kpar^2 - 2 k_z^2)$ & $ v_z k_z + \beta (\kpar^2 - 2 k_z^2)$
\end{tabular}
\end{ruledtabular}
\end{table*}

Because the model Hamiltonian for each node has cylindrical symmetry, the equation of motion for the interband coherence (cf. Eq.~(\ref{eq:P2b})) may be reduced to an effective two-dimensional problem in momentum space, which speeds up its numerical solution very significantly.
To see this, we begin by expanding 
\begin{equation}
\label{eq:Pm}
P_\tau({\bf k}, \omega) = \sum_m e^{i m\varphi} P_{m\tau}(\kpar,k_z, \omega),
\end{equation}
where $\varphi$ is the azimuthal angle of ${\bf k}$, $\kpar=|{\bf\kpar}|$ and $m\in\mathbb{Z}$ is a good (angular-momentum) quantum number associated with the cylindrical symmetry of the model. 
Replacing Eq.~(\ref{eq:Pm}) in Eq.~(\ref{eq:P2b}), multiplying both sides of the resulting equation by $\exp(-i m'\varphi)$, integrating over $\varphi$, and recognizing that $V({\bf k},{\bf k}')$ depends on the azimuthal angles only through $\phi\equiv\varphi-\varphi'$, we obtain
\begin{widetext}
\begin{equation}
\label{eq:P2bm}
\left[\omega+i\delta-2\tildeE\right] P_{m\tau}(\kpar, k_z, \omega) = \Delta f_\tau({\bf k}) \left[\boldsymbol{{\cal E}}(\omega)\cdot{\bf d}_{m\tau}(\kpar,k_z)+ \int'_{\kpar',k_z'} V_{m\tau}(\kpar,k_z; \kpar',k_z') P_{m\tau} (\kpar',k_z', \omega)\right],
\end{equation}
where $2 \tildeE \equiv 2\E+ \dE$ is the renormalized interband transition energy,  
\begin{equation}
\label{eq:vm}
V_{m\tau}(\kpar,k_z; \kpar',k_z') = g  v \frac{1}{2}\int_0^{2\pi}\frac{d\phi}{2\pi} e^{-i m\phi}\frac{\sin\theta_\tau\sin\theta_\tau'+(1+\cos\theta_\tau\cos\theta_\tau')\cos\phi+i(\cos\theta_\tau+\cos\theta_\tau')\sin\phi}{\epsilon({\bf k}-{\bf k}')\left[(k_z-k_z')^2+\kpar^2+\kpar'^2-2\kpar\kpar' \cos\phi\right]}
\end{equation}
\end{widetext}
is the effective Coulomb interaction between the electron and the hole in the $m$-th channel at node $\tau$,
\begin{equation}
g=\frac{e^2}{\epsilon_0\epsilon_\infty  v}
\end{equation} 
is the dimensionless parameter quantifying the strength of Coulomb interactions, and 
\begin{equation}
{\bf d}_{m\tau}(\kpar,k_z) \equiv \int_0^{2\pi} \frac{d\varphi}{2\pi} e^{-i m \varphi} {\bf d}_\tau({\bf k})
\end{equation}
is the interband dipole matrix element projected onto the the $m$-th channel.
To lighten the notation of Eq.~(\ref{eq:P2bm}), we have introduced
\begin{align}
\label{eq:defi}
\Delta f_\tau({\bf k}) &\equiv f_{c\tau}(\kpar,k_z)-f_{v\tau}(\kpar, k_z),\nonumber\\
\int'_{\kpar,k_z} &\equiv  \int\frac{d k_z}{2\pi}\int\frac{d\kpar \kpar}{2\pi} \Theta(\Lambda-\E),
\end{align}
where $\Theta(x)$ is the Heaviside function imposing the ultraviolet cutoff. 

From Eq.~(\ref{eq:vm}), it follows that $V_{m\tau}$ is purely real. 
Similarly, it is easy to see that the difference between $V_{m, \tau}$ and $V_{-m,\tau}$ is proportional to $\cos\theta_\tau+\cos\theta'_\tau$, which can be related to fluxes of the joint particle-hole Berry curvature through surfaces of the type shown in Fig.~\ref{fig:sphere}.

As expected from symmetry, different values of $m$ do not couple in Eq.~(\ref{eq:P2bm}).
Much like for Eq.~(\ref{eq:P2b}), the strategy to solve Eq.~(\ref{eq:P2bm}) is to write
\begin{equation}
P_{m\tau}(\kpar, k_z, \omega) =\sum_n a_{n m \tau}(\omega) \psi_{n m\tau}(\kpar, k_z),
\end{equation}
where $a_{n m}(\omega)$ are coefficients to be determined and $\psi_{n m}(\kpar,k_z)$ is a solution of the Wannier equation
\begin{align}
\label{eq:wan2}
&(2 \tildeE -\epsilon_{n m\tau})\psi_{n m \tau}(\kpar,k_z)\nonumber\\
&=\Delta f_\tau({\bf k}) \int'_{\kpar',k_z'} V_m(\kpar,k_z; \kpar',k_z') \psi_{nm\tau}(\kpar',k_z').
\end{align}
This equation can be recasted in the form of an eigenvalue problem, which implies diagonalizing a real non-symmetric matrix (whose eigenvalues $\epsilon_{n m \tau}$ will nonetheless be real).
Proceeding exactly like in the derivation of Eq.~(\ref{eq:Pk}), we arrive at
\begin{widetext}
\begin{align}
{\bf P}_\tau(\omega)&=\sum_{n m} \frac{1}{\omega+i\delta-\epsilon_{n m\tau}}\int'_{\kpar,k_z} {\bf d}^*_{m\tau}(\kpar,k_z)\psi_{ n m \tau}(\kpar,k_z) \int_{\kpar',k_z'}\Delta f_\tau({\bf k}') \psi^*_{n m \tau}(\kpar',k_z') {\bf d}_{m\tau}(\kpar',k_z')\cdot\boldsymbol{{\cal E}}(\omega)\nonumber\\
&+\sum_{n m} \frac{1}{-\omega-i\delta-\epsilon_{n m\tau}}\int'_{\kpar,k_z} {\bf d}_{m\tau}(\kpar,k_z)\psi^*_{ n m \tau}(\kpar,k_z) \int_{\kpar',k_z'}\Delta f_\tau({\bf k}') \psi_{n m \tau}(\kpar',k_z') {\bf d}^*_{m\tau}(\kpar',k_z')\cdot\boldsymbol{{\cal E}}(\omega),
\end{align}
where ${\bf d}^*_{m\tau}(\kpar,k_z)$ is the complex conjugate of ${\bf d}_{m\tau}(\kpar,k_z)$ and we have used ${\cal E}^*(-\omega)={\cal E}(\omega)$.
From this equation, we extract the valley-resolved susceptibility tensor, which has the block-diagonal form
\begin{equation}
\boldsymbol{\chi}_\tau = \left(\begin{array}{ccc} \chi_{\tau, xx} & \chi_{\tau, x y} & 0\\
                                    -\chi_{\tau, x y} & \chi_{\tau, xx} & 0\\
                                    0              & 0             & \chi_{\tau, zz}
\end{array}\right).
\end{equation}
The $xy$ block can be diagonalized by switching to the basis of RCP and LCP light propagating along $z$: $(E_x,E_y)\to(E_+, E_-)$, where $E_\pm=E_x\pm i E_y$.
The corresponding eigenvalues are $\chi_{\tau \pm}$.
Hereafter, we concentrate on the imaginary parts of these (positive) eigenvalues, denoted $\chi''_{\tau,+}$ and $\chi''_{\tau,-}$, which give the absorption coefficients for RCP and LCP electromagnetic waves whose propagation direction is along $z$, respectively.
In Ref.~[\onlinecite{chan2016}], it has been shown that circularly polarized light leads to a shift in the position of Weyl nodes.
This effect does not take part in our expressions for the {\em linear} susceptibility $\chi_{\tau\pm}$, though it would have to be taken into account in the full solution of the semiconductor Bloch equations. 
For the $\tau=1$ node, some lengthy but straightforward algebra yields
\begin{align}
\label{eq:chipp1}
\chi''_{1, \pm}(\omega) &= -\frac{\pi e^2}{16} \sum_n \delta(\omega-\epsilon_{n, m=\pm 1, \tau=1}) \int'_{\kpar,k_z}\left[ v(1+\alpha k_z)(1\pm\cos\theta_1) \mp 2\beta \kpar \sin\theta_1\right]\frac{\psi_{n, m=\pm 1, \tau=1} (\kpar,k_z)}{|{\bf B}_1(\kpar,k_z)|}\nonumber\\ 
&~~~~~~~~~~~~~~\times\int'_{\kpar',k_z'} \Delta f_1({\bf k}) \left[ v(1+\alpha k_z') (1\pm \cos\theta'_1) \mp 2\beta \kpar' \sin\theta'_1\right]\frac{\psi^*_{n, m=\pm 1, \tau=1} (\kpar',k_z')}{|{\bf B}_1(\kpar',k_z')|},
\end{align}
\end{widetext}
where we have used the fact that $\omega>0$, so that $\delta(\omega+\epsilon_{n m \tau})=0$.
Using Tables \ref{tab:TRS} and \ref{tab:P}, the absorption coefficients for the three other nodes can be readily deduced.
For example, $\chi''_{2,\pm}$ can be obtained from $\chi''_{1,\pm}$ via $v\to -v$, $v_z\to -v_z$, and $\alpha\to-\alpha$.
In the time-reversal-symmetric WSM, $\chi''_{3,\pm}$ ($\chi''_{4,\pm}$) can be obtained from $\chi''_{1,\pm}$ ($\chi''_{2,\pm}$) by $\alpha\to-\alpha$ and $\beta\to-\beta$.
In an inversion-symmetric WSM, $\chi''_{3,\pm}=\chi''_{2,\pm}$ and $\chi''_{4,\pm}=\chi''_{1,\pm}$.
An important observation from Eq.~(\ref{eq:chipp1})  is that only $m=1$ particle-hole excitations contribute to $\chi''_{\tau, +}$ (optical absorption of RCP light), whereas only $m=-1$ particle-hole excitations contribute to $\chi''_{\tau,-}$ (optical absorption of LCP light).
This selection rule is a consequence of taking the wave vector of the light parallel to the wave vector that connects two Weyl nodes of opposite chirality.
It is also the reason why the difference between $V_{m\tau}$ and $V_{-m,\tau}$, alluded to after Eq.~(\ref{eq:defi}), can lead to a different optical absorption for LCP and RCP lights.

One can similarly derive an expression for $\chi''_{\tau, zz}$, which will involve only $m=0$ particle-hole excitations.
Since the most interesting physical effects emerge under circularly polarized light, we will not consider $\chi''_{\tau, zz}$ from here on.


\section{Numerical results}
\label{sec:numerical}

In order to evaluate the optical absorption for LCP and RCP lights, we solve Eq.~(\ref{eq:wan2}) following the numerical approach of Ref.~[\onlinecite{garate2011}], and afterwards enter the solution into Eq.~(\ref{eq:chipp1}) (or variants thereof, in the case of $\tau\neq 1$ nodes).
The Dirac delta function of Eq.~(\ref{eq:chipp1}) is approximated by a gaussian with a standard deviation of $0.6 |\epsilon_F|$. 
In the numerical calculation, we discretize the momenta $k_\parallel$ and $k_z$ into $N=82$ points {\em each}, following a Gauss-Legendre quadrature.
By redoing the calculation with $N=116$, we have verified that the numerical results for the optical absorption have already converged at $N=82$. 
Also, we take $v=2.5 \times 10^5 {\rm m/s}$, $v_z=1.3 v$ and $\epsilon_\infty = 30$ everywhere ($g\simeq 0.6$), except for the case of non-interacting WSM (in which case $\epsilon_\infty \to \infty$).
Concerning the dielectric function $\epsilon({\bf q})$, we adopt the Thomas-Fermi approximation with the screening wave vector $k_s=\sqrt{v g \rho_\tau(\epsilon_F)}$, $\rho_\tau(\epsilon_F)$ being the node-resolved density of states at the Fermi energy (we also add the leading $q\neq 0$ corrections, though they do not make a significant impact).
Finally, unless otherwise stated, we take $\Lambda = 10 |\epsilon_F|$.   

\begin{figure}[t]
\includegraphics[width=0.4\textwidth]{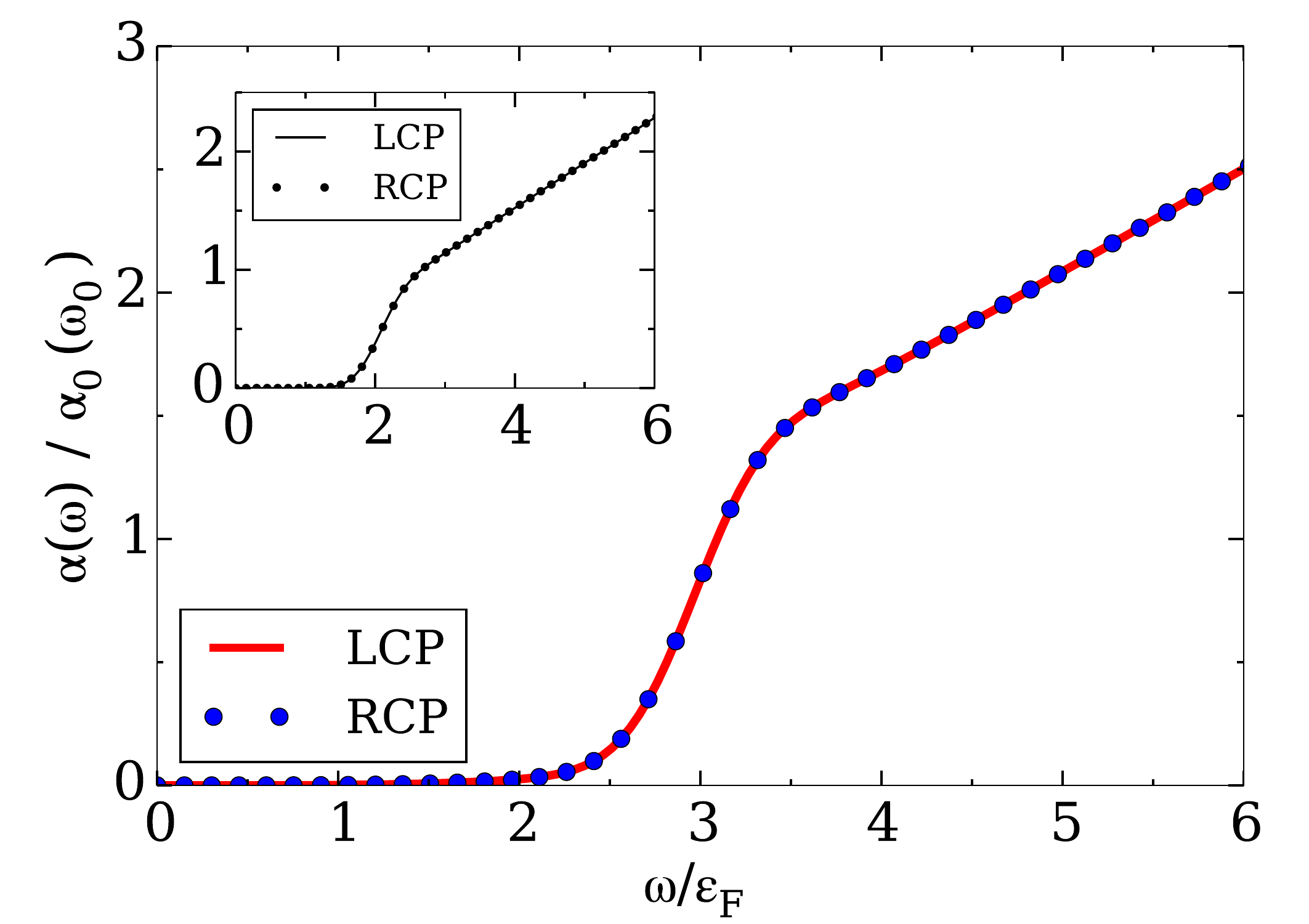}
\caption{
Optical absorption coefficient for the $\tau=1$ node in a Weyl semimetal with a {\em perfectly linear} single-particle energy spectrum and Coulomb interactions.
By definition, $\alpha_0(\omega)$ is the optical absorption coefficient in the absence of Coulomb interactions, and $\omega_0\simeq 2.66 |\epsilon_F|$ is the frequency beyond which $\alpha_0(\omega)$ {\em vs.} $\omega$ becomes linear.
The self-energy contribution shifts the optical absorption threshold from the non-interacting $\omega=2|\epsilon_F|$ to $\omega\simeq 3|\epsilon_F|$. 
The main message from this figure is that the left- and right-circularly polarized lights yield an identical absorption spectrum.
{\em Inset:} optical absorption in the absence of Coulomb interactions. In this case, the absorption coefficient can be calculated analytically. We have verified that the analytical result is in agreement with the numerical one.
}
\label{fig:linear}
\end{figure}

\subsection{Single Weyl node}

Let us first discuss our results for a single Weyl node, e.g. the $\tau=1$ node.
For a Weyl node with perfectly linear dispersion, the optical absorption for RCP and LCP lights turns out to be identical regardless of Coulomb interactions (see Fig.~\ref{fig:linear}).
Mathematically, the RCP-LCP degeneracy originates from the averaging out of the $\cos\theta_\tau+\cos\theta'_\tau$ term in Eq.~(\ref{eq:vm}).
Heuristically, the lack of chirality effects in the optical absorption of a linear WSM can be understood from the arguments sketched in Fig.~\ref{fig:cartoon0}. 
Physically, the degeneracy is a consequence of a pseudo time-reversal symmetry of the linear model,
\begin{equation}
\label{eq:pseudo}
h_\tau (-{\bf k}) = {\cal T}^{-1} h_\tau({\bf k}) {\cal T},
\end{equation}
where ${\cal T}=i \sigma^y K$ and $K$ is the complex conjugate operator.
Under ${\cal T}$, $m \to -m$ and hence RCP $\to $ LCP. 
Thus, if the model Hamiltonian has a pseudo time-reversal symmetry, the absorption coefficient must be the same for RCP and LCP.
This result is at first glance disappointing, because it establishes the degeneracy of LCP and RCP absorption spectra in spite of the nontrivial Berry curvature.

\begin{figure}[t]
\includegraphics[width=0.4\textwidth]{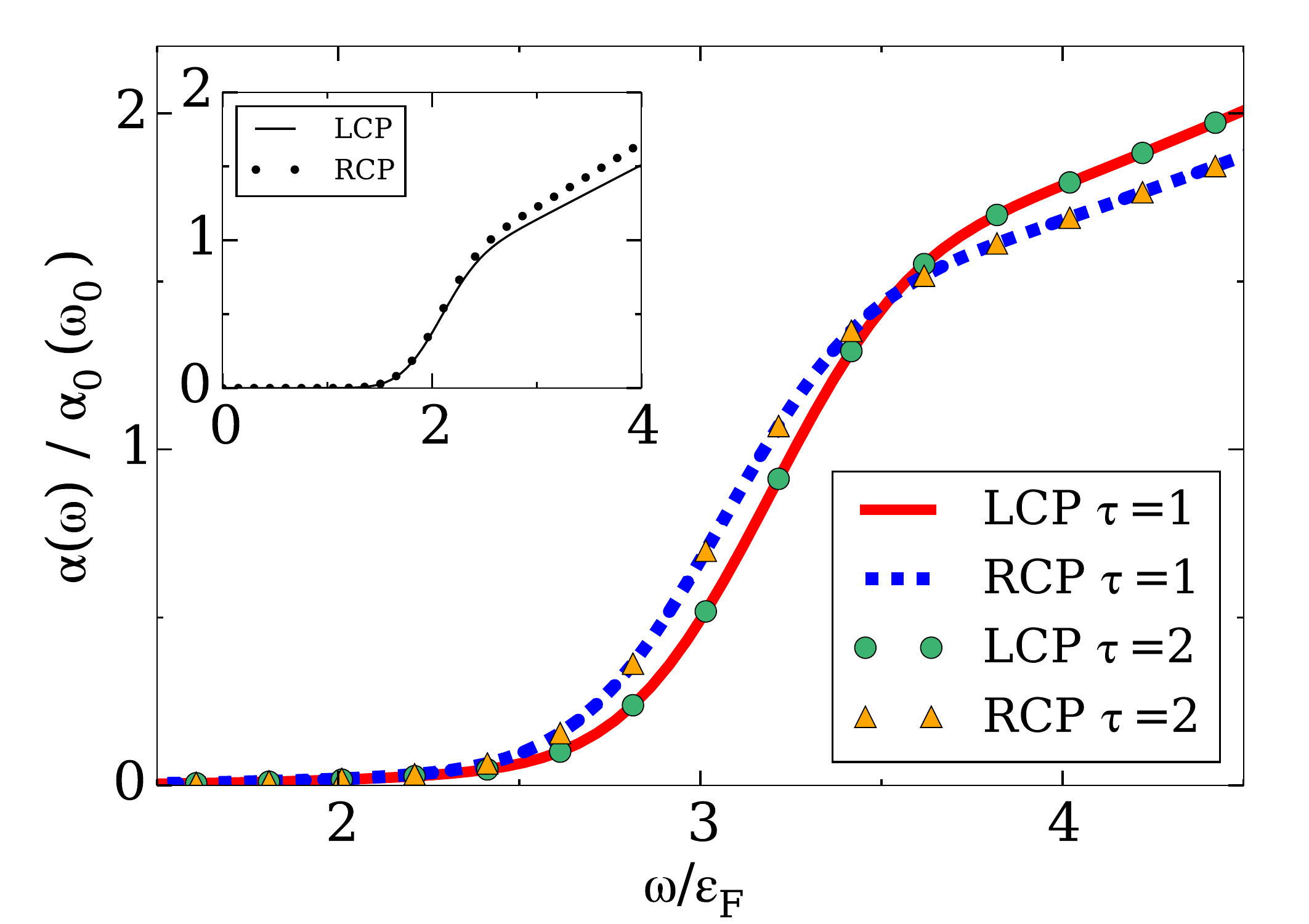}
\caption{Optical absorption of $\tau=1$ and $\tau=2$ nodes (which are mirror partners), in presence of nonlinear terms in the energy spectrum, 
for $\alpha |\epsilon_F|/v=0.075$ and $\beta |\epsilon_F|/v^2=0.02$ (these are rather conservative values for the nonlinear parameters).
The free electron absorption coefficient $\alpha_0(\omega)$ and $\omega_0$ have been defined in the caption of Fig.~\ref{fig:linear}.
For a given node, the absorption spectra for left- and right-circularly polarized lights differ.
For a given handedness of the incident light, the absorption intensity is the same in mirror-related nodes.
{\em Inset:} Optical absorption for the $\tau=1$ node, with the same band parameters as in the main figure, but without Coulomb interactions. In this case, the RCP-LCP asymmetry at the absorption threshold is significantly weaker.
}
\label{fig:nonlin1}
\end{figure}

\begin{figure}[t]
\includegraphics[width=0.4\textwidth]{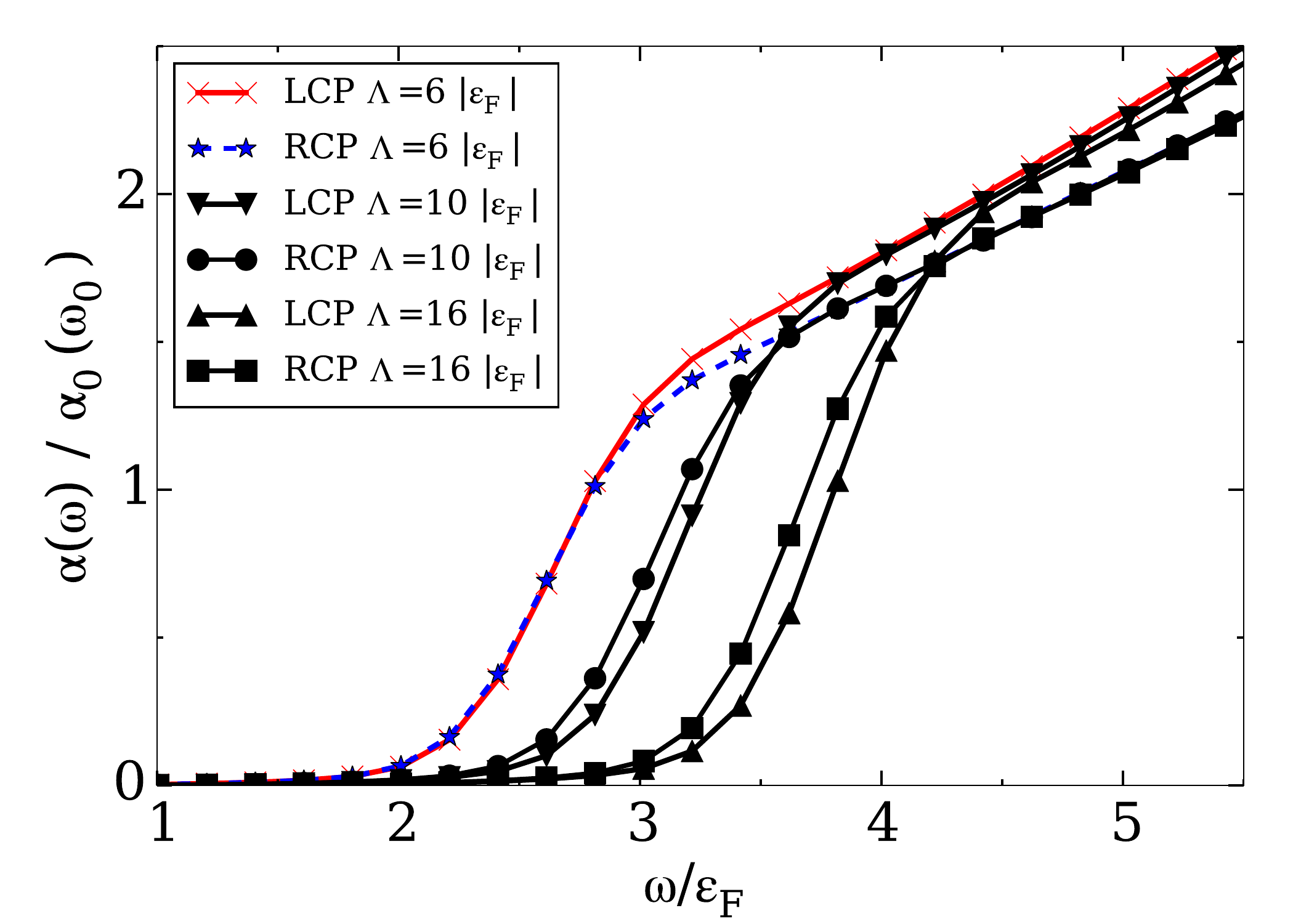}
\caption{Dependence of the optical absorption on $\Lambda$, the ultraviolet cutoff of the model, for $\alpha |\epsilon_F|/v=0.075$ and $\beta |\epsilon_F|/v^2=0.02$.
The free electron absorption coefficient $\alpha_0(\omega)$ and $\omega_0$ have been defined in the caption of Fig.~\ref{fig:linear}. 
A larger cutoff results in a larger difference between the optical absorption of right- and left-circularly polarized lights.
At any rate, a nonzero RCP-LCP difference persists regardless of the value of $\Lambda$, insofar as $\alpha$ or $\beta$ are nonzero.
In particular, the LCP and RCP curves for $\Lambda = 6 |\epsilon_F|$ in this figure differ by about $5\%$ at the optical absorption threshold (the difference then grows gradually at higher photon frequencies).}
\label{fig:nonlin3}
\end{figure}

\begin{figure}[t]
\includegraphics[width=0.4\textwidth]{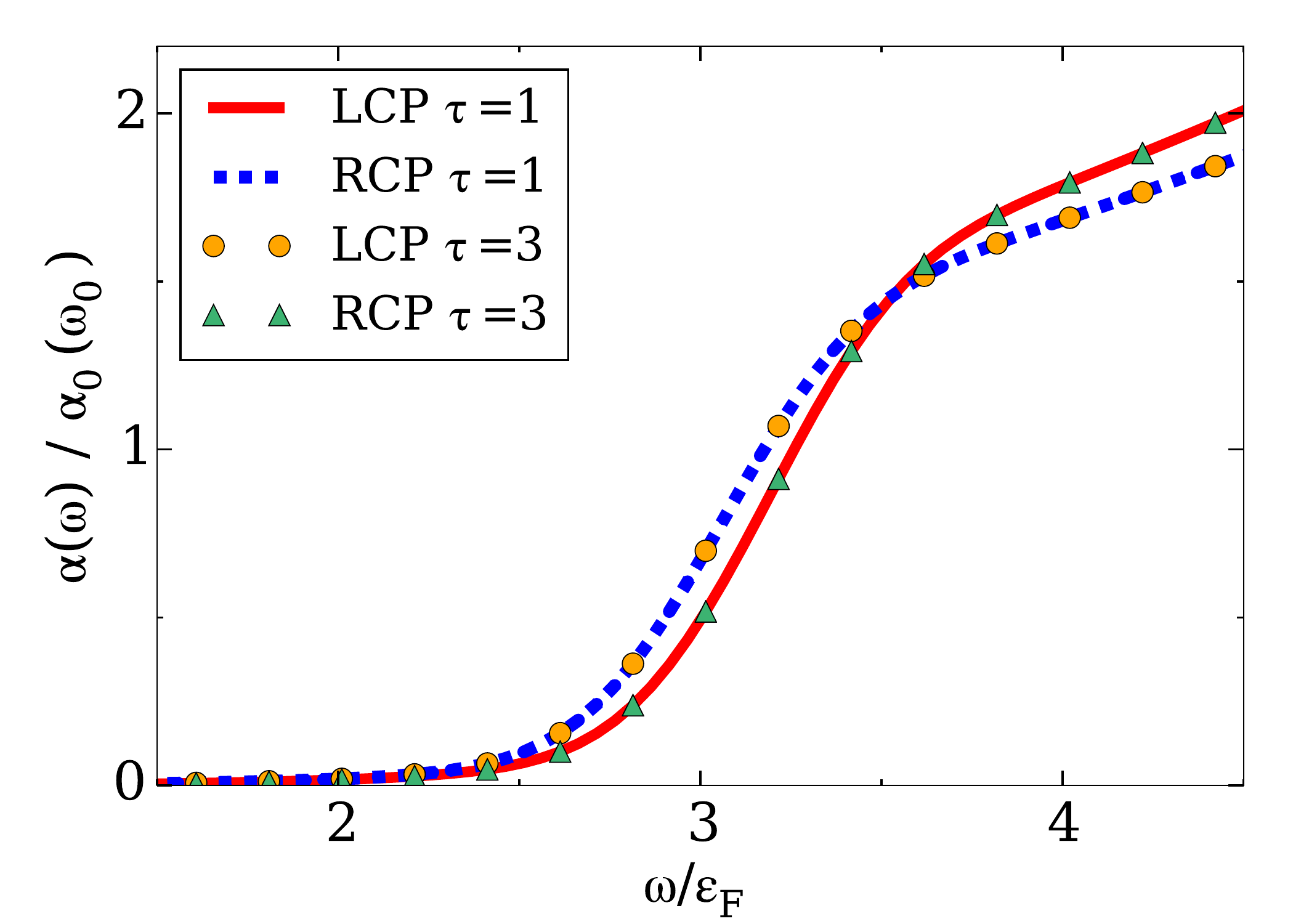}
\caption{
Optical absorption at $\tau=1$ and $\tau=3$ nodes in a Weyl semimetal with time-reversal symmetry, 
for $\alpha |\epsilon_F|/v=0.075$ and $\beta |\epsilon_F|/v^2=0.02$. 
The two nodes are partners under time-reversal. 
The LCP absorption for $\tau=1$ is identical to the RCP absorption for $\tau=3$, and vice versa. 
The free electron absorption coefficient $\alpha_0(\omega)$ and $\omega_0$ have been defined in the caption of Fig.~\ref{fig:linear}.
}
\label{fig:nonlin2}
\end{figure}

However, the situation becomes more interesting when nonlinear terms in the energy spectrum are incorporated. 
These terms break the pseudo time-reversal symmetry, i.e.,  Eq.~(\ref{eq:pseudo}) is no longer obeyed.
Consequently, RCP and LCP lights can, and do, produce different absorption spectra (see Fig.~\ref{fig:nonlin1}).
This is a qualitatively new effect that cannot be captured in the linear approximation.

Excluding self-energy effects, the difference between the LCP and RCP absorption intensities at frequency $\omega$ is governed by the dimensionless parameters 
\begin{equation}
\frac{\alpha \omega}{v} \text { and }   \frac{\beta \omega}{v^2},
\end{equation}
assuming $v_z\simeq v$.
If these dimensionless parameters are small compared to unity, the RCP-LCP splitting is small.
Consequently, the RCP-LCP splitting is larger at higher frequencies of the incident light.
Along the same lines, $\alpha |\epsilon_F|/v$ and $\beta |\epsilon_F|/v^2$ determine the magnitude of the RCP-LCP asymmetry near the optical absorption threshold ($\omega\simeq 2 |\epsilon_F|$).
Hence, one way to enhance the RCP-LCP difference near the threshold is to increase the equilibrium hole concentration of the WSM. 
In addition, we find that the RCP-LCP asymmetry near the threshold is greatly amplified by Coulomb interactions.
This is particularly true for the situation with $\beta=0$: in this case, $\alpha\neq 0$ will not induce any asymmetry between RCP and LCP absorption spectra {\em unless} Coulomb interactions are included.
Finally, whether RCP absorption is stronger or weaker than LCP absorption depends on the details of the Coulomb interactions and the electronic structure, though one important observation is that the RCP-LCP difference changes sign when both $\alpha$ and $\beta$ reverse their signs.



From a numerical standpoint, the RCP-LCP splitting comes from two sources. 
One of the sources is the last term in the numerator of Eq.~(\ref{eq:vm}), containing $\cos\theta_\tau+\cos\theta'_\tau$; this term is no longer averaged-out in the presence of nonlinear terms in the energy spectrum.
The second source of the RCP-LCP difference is the self-energy term in Eq.~(\ref{eq:wan2}).
Although the self-energy is independent of $m$, it produces an anisotropic optical gap in presence of nonlinear terms in the energy spectrum, which 
affects differently the LCP and RCP absorption due to the disparity between the $m=1$ and $m=-1$ dipole matrix elements. 
Moreover, because the self-energy depends on the UV cutoff, it introduces another pair of dimensionless parameters characterizing the RCP-LCP splitting, namely $\alpha \Lambda/v$ and $\beta \Lambda/v^2$.

Unexpectedly, the origin of the RCP-LCP asymmetry does not reside in the difference between particle-hole excitation energies with $m=1$ and $m=-1$; we find these energies to be very similar to each other.
If we ignore self-energy effects, the RCP-LCP asymmetry results purely from the difference between the {\em wave functions} corresponding to $m=1$ and $m=-1$ particle-hole pairs.
We will return to this point in Sec.~\ref{sec:analytical}, where it will be shown that the difference between the $m=1$ and $m=-1$ wave functions can be given a topological interpretation.


For completeness, Fig.~\ref{fig:nonlin3} shows the dependence of the optical absorption on the ultraviolet cutoff of the model.
The main impact of the cutoff on our results takes place via the self-energy term, which shifts the optical absorption threshold to higher frequencies.
A larger cutoff implies a larger self-energy correction. 
It follows that a larger cutoff will produce a larger the RCP-LCP splitting near the (renormalized) absorption threshold, because (i) $\alpha \omega/v$ and $\beta \omega/ v^2$ become larger due to an increased threshold frequency, and (ii) $\alpha \Lambda/v$ and $\beta\Lambda/v^2$ become larger as well.
As a result, the valley polarization near the optical absorption threshold can vary from a few percent to several tens of percent as a function of the cutoff.
Consequently, a quantitative study of the valley polarization in WSM  will require starting from an electronic structure that is devoid of a cutoff.
This task is beyond the scope of the present work.    
At any rate, the qualitative features of the optical absorption spectrum are cutoff-independent.


\begin{figure}[t]
\includegraphics[width=0.4\textwidth]{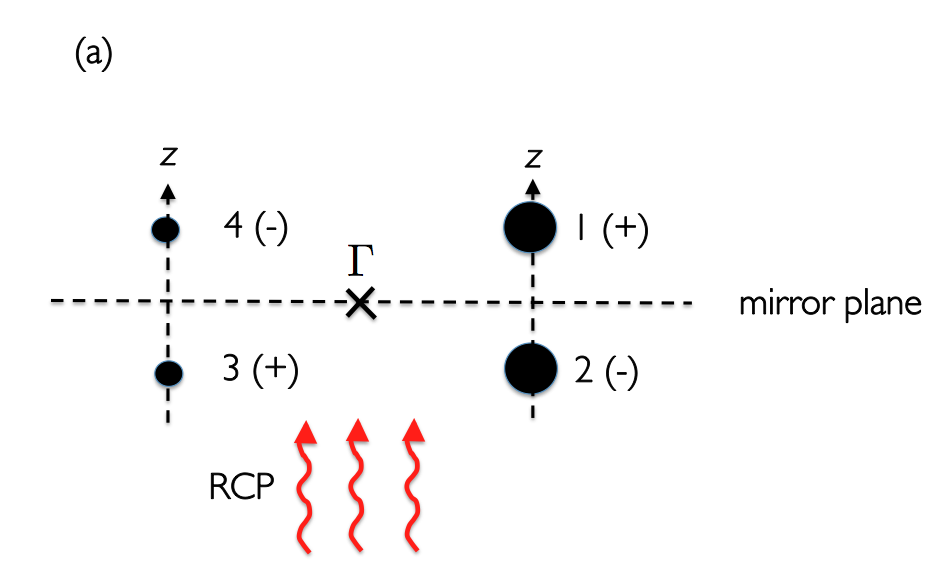}
\includegraphics[width=0.4\textwidth]{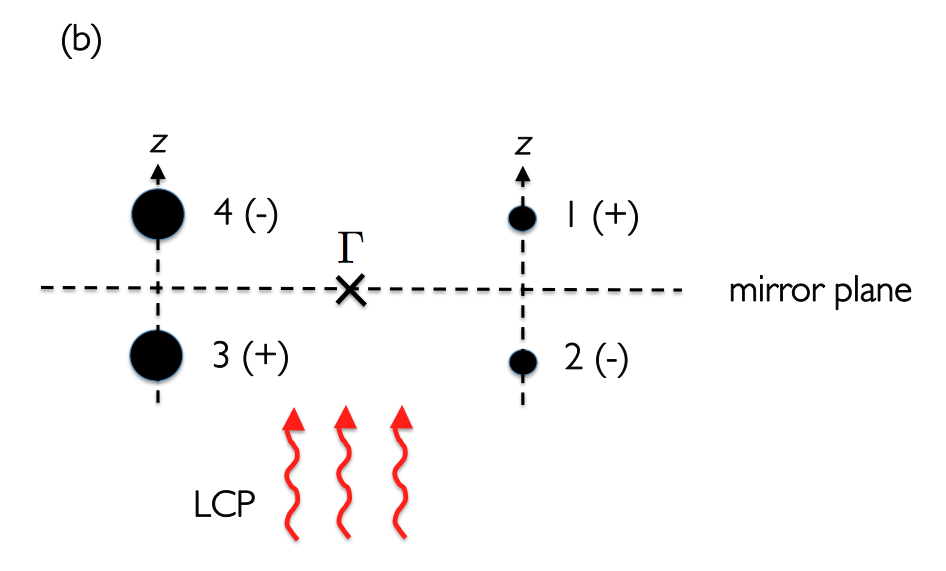}
\caption{Schematic representation of the light-induced valley polarization in a Weyl semimetal with time-reversal and mirror symmetries.
Weyl nodes are labelled with numbers; their respective chiralities are denoted in parenthesis.
The size of the black circles represents the magnitude of the optical absorption coefficient near the absorption threshold at each node.
The $\Gamma$ point is shown for reference purposes. 
(a) When the incident light is right-circularly polarized and has a propagation direction along $z$, nodes 1 and 2 host stronger absorption than nodes 3 and 4 in the vicinity of the absorption threshold. 
(b) When the incident light is left-circularly polarized and has a propagation direction along $z$, nodes 3 and 4 host stronger absorption than nodes 1 and 2 in the vicinity of the absorption threshold.
}
\label{fig:cartoon1}
\end{figure}

\begin{figure}[t]
\includegraphics[width=0.4\textwidth]{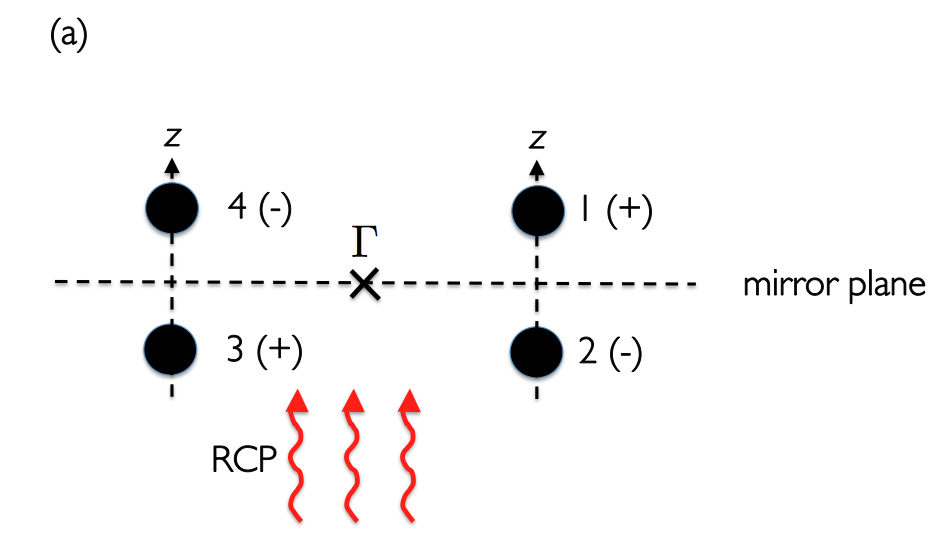}
\includegraphics[width=0.4\textwidth]{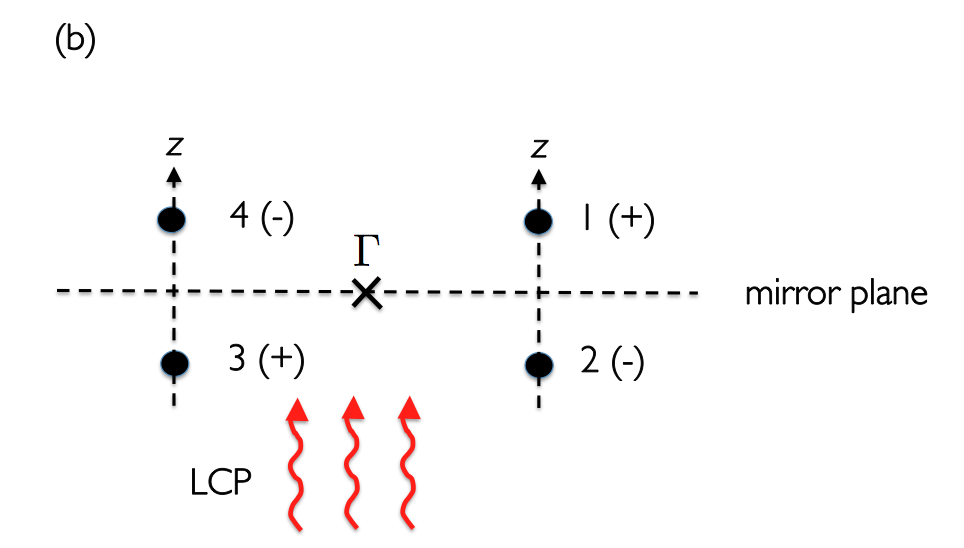}
\caption{Schematic representation of the optical absortion in a Weyl semimetal with inversion and mirror symmetries.
Weyl nodes are labelled with numbers; their respective chiralities are denoted in parenthesis.
The size of the black circles represents the intensity of the optical absorption at each node.
The $\Gamma$ point is shown for reference purposes. 
The total absorption coefficient is different for LCP and RCP lights incident along $z$.
}
\label{fig:cartoon2}
\end{figure}

\subsection{Two nodes related by a mirror plane}

Let us now consider the optical absorption in the $\tau=2$ node.
By construction, this node is related to the $\tau=1$ node by a mirror plane perpendicular to $k_z$.
For a given handedness of the incident light, the optical absorption in the $\tau=2$ node is the same as that of the $\tau=1$ node, irrespective of nonlinearities and Coulomb interactions (see Fig.~\ref{fig:nonlin1}). 
Hence, circularly polarized light does not induce a chiral chemical potential in a WSM containing a mirror symmetry.

\subsection{Time-reversal symmetric WSM}

In Fig.~\ref{fig:nonlin2},  we display the optical absorption coefficient for the $\tau=3$ Weyl node, which is the time-reversed partner of node $\tau=1$.
The absorption spectrum for LCP light in the $\tau=1$ node coincides with the absorption spectrum of the RCP light in the $\tau=3$ node.
This is not surprising, because time-reversal transforms LCP light into RCP light.
The situation in this case is illustrated schematically in Fig.~\ref{fig:cartoon1}.
Although the {\em total} optical absorption is the same for the LCP and RCP lights, the partial (valley-resolved) optical absorption is not.
Due to the combined nonlinear energy spectrum and Coulomb interactions, RCP light excites more electron-hole pairs in $\tau=1$ and $\tau=2$ nodes, whereas LCP light excites more electron-hole pairs in $\tau=3$ and $\tau=4$ nodes.
This implies a pairwise valley polarization induced by circularly polarized light.
The valley polarization is amplified by Coulomb interactions and may be significant near the optical absorption threshold.
Although it has been extensively studied in graphene\cite{rycerz2007} and topologically nontrivial 2D insulators,\cite{dai2012} we are not aware of prior theoretical or experimental reports of valley polarization in WSM.


\subsection{Inversion-symmetric WSM}

In our model of WSM with inversion and mirror symmetry, the node-resolved absorption coefficient is the same in all nodes. 
However, this coefficient differs between LCP and RCP lights. 
Hence, the total optical absorption is different for LCP and RCP incident lights (see Fig.~\ref{fig:cartoon2}). 
Such difference in absorption is allowed in a crystal without time-reversal symmetry.
Once again, we emphasize that this effect would be absent in the linear approximation of the energy spectrum around the Weyl nodes. 

\subsection{Reversal of the direction of light propagation}

Thus far, we have assumed that the direction of light propagation is along the positive $z$ direction.
If the direction of propagation is reversed, the roles of LCP and RCP are exchanged\cite{yu2016} and consequently the valley polarization is reversed.
In other words, LCP and RCP are exchanged in Figs.~\ref{fig:nonlin1} and \ref{fig:nonlin2}, while the small and large black circles are exchanged in Figs.~\ref{fig:cartoon1} and \ref{fig:cartoon2}.

\section{Analytical results}
\label{sec:analytical}

The objective of this section is to support and supplement the numerical results of the preceding section with a simplified analytical solution of Eq.~(\ref{eq:wan2}).
Our approach is partly related to that of Ref.~[\onlinecite{shytov2015}], which studied two-electron bound states. 
The main simplification consists of replacing the screened Coulomb potential in real space by a delta function potential. 
This approximation is valid at length scales that far exceed the screening length, i.e., for momenta that are small compared to the Thomas-Fermi screening wave vector $k_s$.
If $k_s$ is large compared to the momentum cutoff of the model (which is mathematically possible in the large $g$ limit, or in the high-doping limit, or else in the neighborhood of a van-Hove singularity for the density of states), but still small compared to the separation between the Weyl nodes, then we can approximate Eq.~(\ref{eq:vm}) as
\begin{widetext}
\begin{align}
\label{eq:vm2}
V_{m\tau}(\kpar,k_z; \kpar',k_z') &\simeq \frac{g v}{2 k_s^2}\int_0^{2\pi}\frac{d\phi}{2\pi} e^{-i m\phi}\left[\sin\theta_\tau\sin\theta_\tau'+(1+\cos\theta_\tau\cos\theta_\tau')\cos\phi+i(\cos\theta_\tau+\cos\theta_\tau')\sin\phi\right]\nonumber\\
&=\frac{g v}{2 k_s^2}\left[\sin\theta_\tau \sin\theta'_\tau\delta_{m,0} + \frac{1}{2} (1-\cos\theta_\tau)(1-\cos\theta_\tau')\delta_{m,-1} + \frac{1}{2}(1+\cos\theta_\tau)(1+\cos\theta_\tau')\delta_{m,1}\right].
\end{align}
\end{widetext}
In this approximation, only $m=0, \pm 1$ channels contribute to the effective electron-hole attraction.
Out of these, only the $m=\pm 1$ are active under irradiation by LCP and RCP lights.
In addition, Eq.~(\ref{eq:vm2}) becomes independent of the interaction strength because $g/k_s^2$ is independent of $g$ in the Thomas-Fermi approximation.
Finally, the interaction kernel is separable into ``primed'' and ``non-primed'' variables, which will enable an analytical solution of the corresponding Wannier equation.
In fact, the problem at hand becomes a variation of the Cooper problem in the BCS theory of superconductivity.\cite{schrieffer1957}

Let us consider the $m=0$ channel first.
Dividing both parts of Eq.~(\ref{eq:wan2}) by $(2\E-\epsilon_{n,m=0,\tau})$ (which we assume to be nonzero), multiplying by $\sin\theta_\tau$ and integrating over ${\bf k}$, we arrive at the condition
\begin{equation} 
\label{eq:e0}
\frac{g v}{2 k_s^2} \int'_{\kpar,k_z}\sin^2\theta_\tau \frac{\Theta(\E-|\epsilon_F|)}{2\E-\epsilon_{n,m=0,\tau}}=1,
\end{equation}
where we have taken the zero temperature limit, and $\epsilon_F<0$ (hole-doped WSM).
Besides, for simplicity, we have neglected the self-energy correction to the energy bands, so that $\tildeE\to\E $.
We remind the reader that the integrals over momenta are constrained by the condition $\E<\Lambda$ (cf. Eq.~(\ref{eq:defi})).

Equation(\ref{eq:e0}) gives the electron-hole excitation energies corresponding to $m=0$, at the valley $\tau$.
Proceeding in the same way, we find that the excitation energies for the $m=\pm 1$ channels must obey
\begin{equation}
\label{eq:epm1}
\frac{ v g}{ 2 k_s^2}\int'_{\kpar,k_z}\frac{(1\pm \cos\theta_\tau)^2}{2}\frac{\Theta(\E-|\epsilon_F|)}{2\E-\epsilon_{n,m=\pm 1,\tau}}=1.
\end{equation}
In order to obtain approximate analytical solutions of Eqs.~(\ref{eq:e0}) and (\ref{eq:epm1}),
 we begin by recognizing that
\begin{equation}
\label{eq:dos}
\int'_{\kpar,k_z} F(\kpar,k_z)= \int^\Lambda dE \int'_{\bf k} F({\bf k}) \delta (E-|{\bf B}_\tau({\bf k})|),
\end{equation}
where $\int'_{\bf k} \equiv \int d^3 k/(2\pi)^3 \Theta( \Lambda-|{\bf B}_\tau({\bf k})|)$.
Applying Eq.~(\ref{eq:dos}) to Eqs.~(\ref{eq:e0}) and (\ref{eq:epm1}), the latter become   
\begin{align}
\label{eq:eigen}
&\frac{g v}{2 k_s^2} \int_{|\epsilon_F|}^\Lambda dE  \frac{\rho_\tau(E)}{2 E-\epsilon_{n,m=0,\tau}} \langle \sin^2\theta_\tau \rangle_E =1 \nonumber\\
&\frac{g v}{2 k_s^2} \int_{|\epsilon_F|}^\Lambda dE \frac{\rho_\tau(E)}{2 E-\epsilon_{n,m={\pm 1},\tau} } \Big\langle \frac{(1\pm \cos\theta_\tau)^2}{2} \Big\rangle_E =1,
\end{align}
where $\rho_\tau(E)=\int'_{\bf k} \delta(E-|{\bf B}_\tau({\bf k}|)$ is the valley-resolved density of states at energy $E$ and
\begin{equation}
\langle f_\tau ({\bf k}) \rangle_E \equiv \frac{\int'_{\bf k} f({\bf k}) \delta(E-\E)}{\rho_\tau(E)}
\end{equation}
is the average of a function $f$ over a constant energy ($E$) surface in momentum space around the node $\tau$.

The solutions of Eq.~(\ref{eq:eigen}), labeled by the index $n$, are multiple.
Here, we are interested in the solutions of energy $\lesssim 2 |\epsilon_F|$ near the optical absorption threshold.
In this case, the integrands in Eq.~(\ref{eq:eigen}) will be peaked near $E\simeq|\epsilon_F|$ and therefore we arrive at
\begin{align}
\label{eq:sol}
\epsilon_{m=0,\tau} &\simeq 2|\epsilon_F| - 2 \Lambda \exp\left[-1/\lambda_{m=0,\tau}(|\epsilon_F|)\right]\nonumber\\
\epsilon_{m={\pm 1},\tau} &\simeq 2|\epsilon_F| - 2 \Lambda \exp\left[-1/\lambda_{m=\pm 1,\tau}(|\epsilon_F|)\right],
\end{align}
where
\begin{align}
\label{eq:lambdas}
\lambda_{m=0,\tau} (|\epsilon_F|) & =  \frac{g  v}{4 k_s^2} \rho_\tau(|\epsilon_F|)  \langle \sin^2\theta_\tau \rangle_{|\epsilon_F|}\nonumber\\
\lambda_{m=\pm 1,\tau}(|\epsilon_F|) & = \frac{g  v}{8 k_s^2} \rho_\tau(|\epsilon_F|)  \langle (1\pm \cos\theta_\tau)^2 \rangle_{|\epsilon_F|}.
\end{align}
In the derivation of Eq.~(\ref{eq:sol}), we have neglected $O(|\epsilon_F|/\Lambda)$ terms. 
The quantities $2|\epsilon_F|-\epsilon_{m,\tau}$ are the binding energies of Mahan-like excitons\cite{mahan1966} with azimuthal angular momentum $m$.
Also, Eq.~(\ref{eq:sol}) is valid only for exponentially small binding energies ($\lambda_{m\tau}(|\epsilon_F|)\ll 1$).


Let us discuss Eq.~(\ref{eq:sol}) for some special cases.
When $\alpha=\beta=0$ (linear WSM), we find $\langle \cos\theta_\tau\rangle_E=0$ and
\begin{equation}
\langle \sin^2\theta_\tau \rangle_E = \Big\langle \frac{(1\pm \cos\theta_\tau)^2}{2} \Big\rangle_E,
\end{equation}
which implies $\epsilon_{m=0,\tau}=\epsilon_{m=\pm 1,\tau}$. 
Hence, the exciton binding energies in a linear WSM are non chiral.\cite{comment}
 
Next, let us allow for nonlinear terms in the energy dispersion.
It follows that $\langle \cos\theta_\tau\rangle_E\neq 0$. 
To be quantitative, it is convenient to proceed with the following change of variables, 
\begin{equation}
\label{eq:trans}
\int'_{\bf k} = \int'_{{\bf B}_\tau}\Big|\frac{\partial {\bf k}}{\partial {\bf B}_\tau}\Big|,
\end{equation} 
where $\int'_{{\bf B}_\tau} \equiv \int d^3 B_\tau /(2\pi)^3 \Theta(\Lambda-B_\tau)$ and $|\partial {\bf k}/\partial {\bf B}_\tau|$ is the determinant of the Jacobian.
In spherical coordinates,  ${\bf B}_\tau = B_\tau (\sin\theta_\tau \cos\varphi, \sin\theta_\tau \sin\varphi,\cos\theta_\tau)$, with $B_\tau\in [0,\infty)$, $\theta_\tau\in[0,\pi]$ and $\varphi\in[0,2\pi]$. 
The UV cutoff puts a constraint on $B_\tau$, but not in $\theta_\tau$ and $\varphi$; this is one advantage of the coordinate transformation in Eq.~(\ref{eq:trans}).  
The Jacobian is simple only in the case $\beta=0$, which we adopt hereafter.
For instance, in the $\tau=1$ node,
\begin{equation}
\label{eq:jac}
\Big|\frac{\partial {\bf k}}{\partial {\bf B}_1}\Big|= \Big|\frac{v_z}{v^2(v_z+\alpha B_{1,z})^2}\Big|.
\end{equation}
As mentioned above, we take the UV cutoff in such a way that the nonlinear terms are always smaller than the linear terms.
This imposes $\alpha |k_z| <1$, which in turn ensures that $v_z+\alpha B_{1,z} >0$.
Using Eqs.~(\ref{eq:trans}) and (\ref{eq:jac}), we obtain 
\begin{equation}
\label{eq:ks2}
k_s^2 = \frac{e^2}{\epsilon_0\epsilon_\infty} \rho_\tau(\epsilon_F)=\frac{2 g \epsilon_F^2}{\pi} \frac{v_z/v}{v_z^2-\epsilon_F^2 \alpha^2}
\end{equation}
for $\beta=0$.
The presence of a UV cutoff guarantees that $|\alpha|<| v_z/\epsilon_F|$.
Substituting Eq.~(\ref{eq:ks2}) in Eq.~(\ref{eq:lambdas}) and evaluating the integrals in the latter, we arrive at
\begin{align}
\label{eq:lambdas2}
\lambda_{m=\pm 1,\tau=1} (x) &= \frac{1\mp x}{16\pi x^2}\left[1 + \frac{1}{2}\left(\frac{1}{x}-x\right)\ln\frac{1-x}{1+x}\right]\nonumber\\
\lambda_{m=0,\tau=1} (x) &= \frac{1-x}{8\pi x} \left(-1 + \frac{1}{2x} \ln\frac{1+x}{1-x}\right),
\end{align}
where we have once again taken $\beta=0$ and we have defined
\begin{equation}
x\equiv \frac{\alpha |\epsilon_F|}{ v_z}
\end{equation}
as a dimensionless parameter that quantifies the nonlinearities in the single-particle energy spectrum.

Let us analyze some limiting cases of Eq.~(\ref{eq:lambdas2}). 
When $x\ll 1$ (weakly nonlinear regime), we have
\begin{align}
\lambda_{m=0,\tau=1} (x) & \simeq 1/(24\pi) +O (x^2)\nonumber\\
\lambda_{m=\pm 1,\tau=1}(x) & \simeq \left(1\mp x\right)/(24\pi) +O (x^2),
\end{align}
which is clearly compatible with the starting assumption of $\epsilon_{m}\simeq 2|\epsilon_F|$.
Thus, 
\begin{align}
&\epsilon_{m=-1}<\epsilon_{m=0}<\epsilon_{m=1}\,\,\,\, (\text{if $\alpha>0$})\nonumber\\
&\epsilon_{m=1}<\epsilon_{m=0}<\epsilon_{m=-1}\,\,\,\, (\text{if $\alpha<0$}),
\end{align}
i.e $\alpha\neq 0$ creates a chirality ($\epsilon_m\neq \epsilon_{-m}$) in the exciton binding energies at a single Weyl node.

If $x\simeq 1$ (strongly nonlinear regime with $\alpha>0$), we find
\begin{align}
\lambda_{m=0,\tau=1} (x) & \simeq \lambda_{m=1,\tau=1} (x)\simeq 0\nonumber\\
\lambda_{m=-1,\tau=1}(x) & \simeq 1/(8\pi), 
\end{align}
where we have omitted $O(1-x)$ and $O((1-x) \ln(1-x))$ terms.
Similarly, if $x\simeq -1$ (strongly nonlinear regime with $\alpha<0$), we find
\begin{align}
\lambda_{m=0,\tau=1} (x) & \simeq\lambda_{m=-1,\tau=1} (x)\simeq  0\nonumber\\
\lambda_{m=1,\tau=1} (x) & \simeq 1/(8\pi).
\end{align}
In sum, in the large $|\alpha|$ regime, the effect of chirality in the exciton bidning energies becomes more pronounced.
Yet, much like in the weak $\alpha$ regime, the strongest binding for $\alpha>0$ ($\alpha<0$) takes place in the $m=-1$ ($m=1$) channel.
In addition, the results in the strong $\alpha$ regime remain consistent with our starting assumption of $\epsilon_m\simeq 2|\epsilon_F|$.

From Eq.~(\ref{eq:lambdas}), it is clear that the difference between $\epsilon_{m=1,\tau}$ and $\epsilon_{m=-1,\tau}$ originates from $\langle \cos\theta_\tau\rangle_{|\epsilon_F|}\neq 0$.
As mentioned in Fig.~\ref{fig:sphere},  $\cos\theta_\tau$ can be linked to the flux of the joint Berry curvature. 
It is likewise useful to notice that
\begin{equation}
\langle\cos\theta_\tau\rangle_{\epsilon_F} \neq 0 \leftrightarrow \langle \boldsymbol{\Omega}_{\rm Sch}\cdot\hat{\bf z}\rangle_{\epsilon_F} \neq 0,
\end{equation}
where ${\boldsymbol\Omega}_{\rm Sch}$ is the Berry curvature defined in Eq.~(\ref{eq:berrycurv}).
In other words, the projection of the Berry curvature along the direction that connects two Weyl nodes of opposite chirality must have a nonzero average over the Fermi surface in order to produce an asymmetry between $m$ and $-m$ exciton states.
In the linear model ($\alpha=\beta=0$), $\langle \cos\theta_\tau \rangle_{|\epsilon_F|}=0$ and the effect of the Berry curvature in the energy splitting between $m$ and $-m$ pairs averages out.
This is a manifestation of the pseudo time-reversal symmetry of a Weyl node with linear dispersion (cf. Eq.~(\ref{eq:pseudo})).
In presence of nonlinear terms, $\langle \cos\theta_\tau \rangle_{|\epsilon_F|}\neq 0$ and the Berry curvature produces a chirality in the optical absorption.

\begin{figure*}[t]
\includegraphics[width=0.9\textwidth]{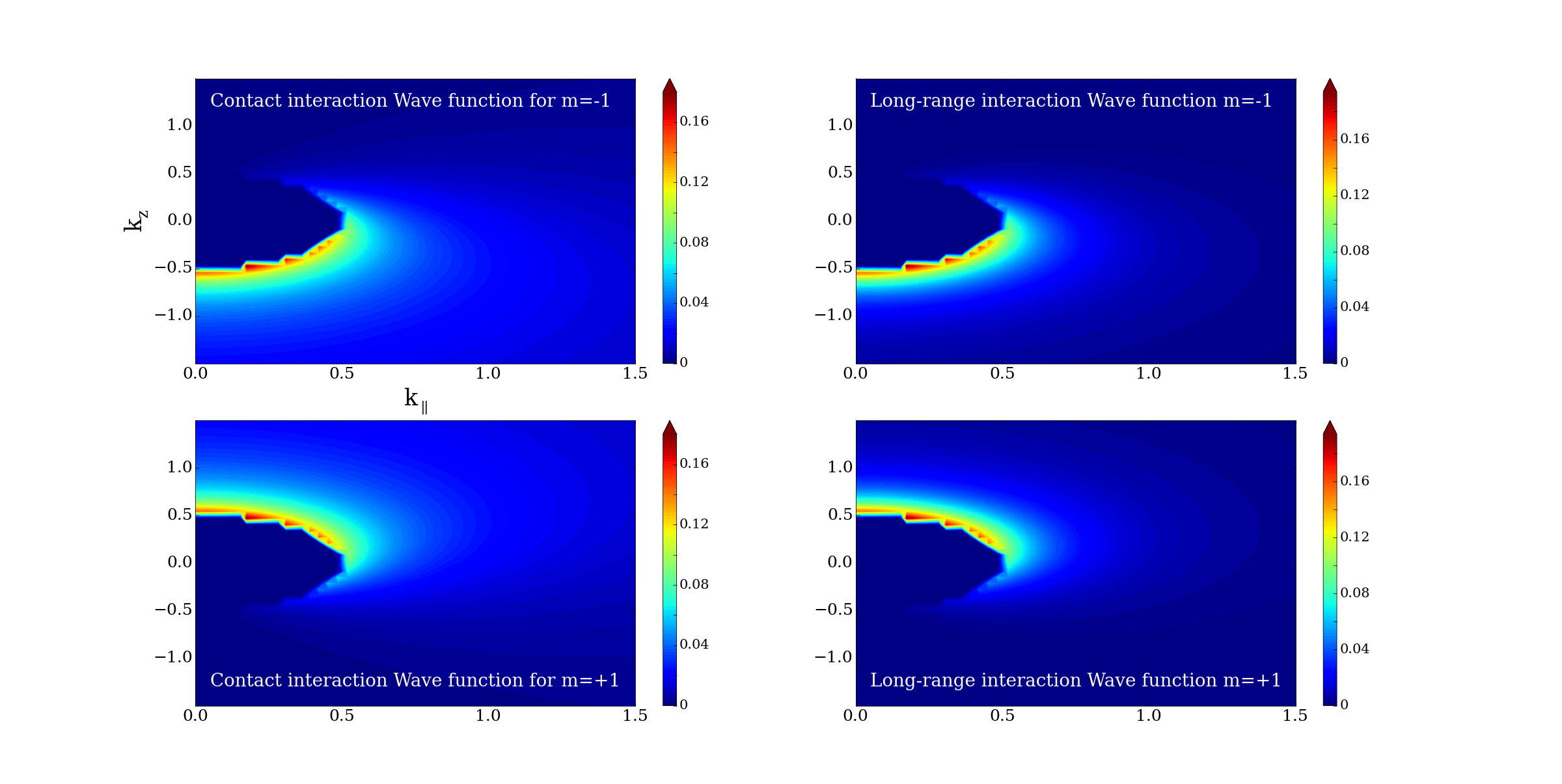}
\caption{
Wave functions for the electron-hole pairs. 
{\em Left pannels:} plots of Eq. (65) for $m =-1$ (top) and $m = 1$ (bottom) wave functions, in the contact interaction model. 
The $m =-1$ wavefunction vanishes when $k_\parallel = 0$ and $k_z > 0$, i.e. when $\cos\theta = 1$. 
In contrast, the $m =1$ wave function vanishes when $k_\parallel = 0$ and $k_z < 0$, i.e. when $\cos\theta = −1$.
{\em Right pannels:} numerically calculated $m = −1$ (top) and $m = 1$ (bottom) wave functions, in the full problem with long-range Coulomb interactions.
Although the details of these wave functions differ with respect to those in the left pannels, they still have nodes at $\cos\theta = \pm 1$ in the case of $m =\mp 1$.
The numerical calculations were done for the linear model ($\alpha=\beta=0$), with $v_z=v$, an energy cutoff $\Lambda =10 |\epsilon_F|$, and the plotted wave function corresponds to  an energy eigenvalue of $2.9607|\epsilon_F|$ (which corresponds to the absorption threshold region for the given value of $\Lambda$). 
Adding non linear terms will lead to different amplitudes for $m=1$ and $m=-1$ wave functions, but will not change the fact that they vanish at $\cos\theta=-1$ and $\cos\theta=1$ (respectively).}
\label{fig:haldane}
\end{figure*}

Thus, our simple analytical model predicts an asymmetry between $m$ and $-m$. 
However, both for $|x|\ll 1$ and $|x|\lesssim 1$, the difference between  $\epsilon_{m=1}$ and $\epsilon_{m=-1}$ is extremely small ($\exp(-24 \pi)\simeq 10^{-33}$ and $\exp(-8\pi)\simeq 10^{-11}$).
In consequence, any realistic broadening of the particle-hole excitation energies will make such difference utterly inconsequential for the optical absorption. 
This conclusion 
is in agreement with the numerical results obtained in Sec.~\ref{sec:numerical} for the full model. 
There, we noted that the difference between $\epsilon_{n m\tau}$ and $\epsilon_{n,-m,\tau}$ is very small for the full model, and that the origin of the asymmetry in the optical absorption lies in the {\em wave functions} of the Wannier equation (when ignoring the self-energy term).
In order to explain this finding, we will now concentrate on the eigenfunctions of Eq.~(\ref{eq:wan2}).
In the approximation of the delta-function interaction, we get
\begin{widetext}
\begin{align}
\label{eq:wavefu}
\psi_{n,m=0,\tau}(\kpar, k_z) &=c_{n,m=0,\tau} \frac{\Theta(\E-|\epsilon_F|)}{2\E-\epsilon_{n, m=0,\tau}} \sin\theta_\tau\nonumber\\
\psi_{n,m=\pm 1,\tau}(\kpar, k_z) &= c_{n, m=\pm 1,\tau}\frac{\Theta(\E-|\epsilon_F|)}{2\E-\epsilon_{n,m=\pm 1,\tau}} (1\pm \cos\theta_\tau),
\end{align}
\end{widetext}
where $c_{nm\tau}$ are constants (independent of ${\bf k}$) that may be determined from the normalization of the wave functions.
The validity of Eq.~(\ref{eq:wavefu}) may be checked by plugging it back into Eq.~(\ref{eq:wan2}), with $V_{m\tau}$ given by Eq.~(\ref{eq:vm2}).

Equation (\ref{eq:wavefu}) shows two features that hold regardless of the presence or absence of nonlinear terms in the single-particle energy spectrum.
First, the particle-hole wavefunctions have nodes occuring at $\theta=0$ for $m=-1$, at $\theta=\pi$ for $m=1$, and at both values of $\theta$ for $m=0$.
Second, the quantum number $m$ gives the vorticity of the wave functions  $\exp(i m \varphi) \psi_{n m \tau} (\kpar, k_z)$  along infinitesimal loops centered on the nodes.
In a way, Eq.~(\ref{eq:wavefu}) is the particle-hole analogue of the topological nodal Cooper pairs proposed by Li and Haldane in superconducting Weyl semimetals.\cite{li2015}
One important difference is, however, that in our case exciton condensation is not necessary in order to have $\psi_{n m \tau}\neq 0$.
 
Figure \ref{fig:haldane} illustrates the wave functions for $m=\pm 1$ as a function of $\kpar$ and $k_z$, thereby confirming the presence of nodes at $\cos\theta=\pm 1$.
Importantly, the same figure shows that the wave functions for the full problem with long range Coulomb interactions also contain nodes at $\cos\theta=\pm 1$.
Consequently, the nodes of the wave functions and their vorticity are topologically robust (i.e.,  independent of the detailed nature of the Coulomb interaction).

Armed with Eq.~(\ref{eq:wavefu}), we can understand analytically why the optical absorption at a given node is different for LCP and RCP.
Starting from Eq.~(\ref{eq:chipp1}), using Eqs.~(\ref{eq:dos}) and (\ref{eq:wavefu}), assuming zero temperature, and (for simplicity) taking $\beta=0$, we arrive at 
\begin{widetext}
\begin{align}
\label{eq:chipp1b}
\chi''_{1,\pm}(\omega) &= -\frac{\pi e^2}{16} \sum_n c^2_{n, m=\pm 1,\tau=1} \delta(\omega-\epsilon_{n, m=\pm 1, \tau=1})\left[\int_{|\epsilon_F|}^\Lambda \frac{dE\, \rho_1(E)}{2 E-\epsilon_{n,m=\pm 1,\tau=1}}\frac{ v}{E}\langle (1+\alpha k_z) (1\pm\cos\theta_1)^2\rangle_E\right]^2
\end{align}
\end{widetext}
We are interested in the values of $n$ such that $\epsilon_{n m \tau}\simeq 2 |\epsilon_F|$. 
Then, as shown above, the particle-hole excitation energies are essentially the same for $m=1$ and $m=-1$.
Likewise, in the linear Weyl model, $\langle (1+\cos\theta)^2 (1+\alpha k_z)\rangle_E = \langle (1-\cos\theta)^2 (1+\alpha k_z)\rangle_E$, and thus $\chi''_{\tau,+} = \chi''_{\tau,-}$.
However, when $\alpha\neq 0$, there is a difference between $\chi''_{\tau,+}$ and $\chi''_{\tau,-}$, which once again originates from a nonzero weighted angular average of $\cos\theta$ and $\cos\theta(1+\alpha k_z)$.
This difference, controlled by the parameter $x$,  does {\em not} involve any exponentially small numbers, and gives the analytical confirmation of the RCP-LCP asymmetry found numerically in Sec.~\ref{sec:numerical}.
An approximate analytical evaluation combining Eqs.~(\ref{eq:wavefu}) and (\ref{eq:chipp1b}) yields $\chi''_{1,+}/\chi''_{1,-} \simeq 1 + x^3/10$ for $x\ll 1$, i.e. RCP absorption is stronger than LCP absorption.
We have verified that this trend is in agreement with the numerical result in the appropriate situation (contact interaction, $\beta=0$, no self-energy term).

We end this section by extending the analytical results to $\tau\neq 1$ nodes.
Let us start with the $\tau=2$ node, which is a mirror partner of the $\tau=1$ node.
In order to transfer the result for $\tau=1$ to $\tau=2$, we apply $v_z\to -v_z$, $v\to -v$, and $\alpha\to -\alpha$.
Clearly, $\epsilon_{n m, \tau=1}=\epsilon_{n m, \tau=-1}$.
Similarly, it can be shown that $\langle (1 + \alpha k_z) (1\pm \cos\theta)\rangle_E$ is the same for the two mirror-related nodes: the key is to notice that $k_z = B_z/v_z$ for $\tau=1$, while $k_z = -B_z/v_z$ for the $\tau=2$ node. 
This gives an analytical explanation to why the optical absorption for mirror-partner Weyl nodes is the same regardless of nonlinearity and interactions.

In a WSM with time-reversal symmetry, the $\tau=3$ node is related to the $\tau=1$ node via $\alpha\to-\alpha$.
Thus, the parameter $x$ changes sign from one node to another.
In this section, we have found analytically that the contribution from nonlinear terms to the optical absorption is an odd function of $x$.
Consequently, we have $\chi''_{\pm,1}=\chi''_{\mp,3}$, which is what we found numerically in Sec.~\ref{sec:numerical}.


\section{Conclusions}

We have presented a theory of the optical absorption for three dimensional Weyl semimetals with a nonlinear energy dispersion, in presence of Coulomb interactions.
The main prediction of this paper is that the node-resolved optical absorption coefficients for right- and left-circularly polarized lights differ, thereby giving rise to a valley polarization.
This effect, whose origin we trace to a nonzero average of the Berry curvature over the Fermi surface, is amplified by Coulomb interactions and emerges only when the nonlinearities in the spectrum are included in the theory.
Thus, it constitutes an example of new physical effects that can arise from the interplay between nontrivial band topology, electron-electron interactions and band curvature in Weyl semimetals.

We have corroborated the preceding numerical results by performing an analytical study of a simple model, where the screened Coulomb interaction is approximated by a contact interaction.
This analytical approach has allowed us to identify electron-hole pairs with exponentially weak binding energies near the optical absorption threshold. 
These particle-hole pairs (generally known as Mahan excitons) turn out to be topologically nontrivial because their wave functions have nodes with nonzero vorticity. 
Due to optical selection rules, left- and right-circularly polarized lights are absorbed by particle-hole pairs with opposite vorticity.
This disparity is in part responsible for the predicted asymmetry in the absorption spectra for left- and right-circularly polarized lights.

The present work can be refined and extended in various ways. 
For example, one can redo the calculation for more general electronic dispersions (tilted Weyl cones, type II Weyl semimetals, dispersions without cylindrical symmetry axis, etc), removing the UV cutoff and incorporating all possible scattering processes due to the Coulomb interaction. 
This would enable a quantitative study of the valley polarization predicted in this work.
In addition, it would be interesting to study the impact of static magnetic fields in our results.

We acknowledge financial support from Qu\'ebec's RQMP, Canada's NSERC, and the Canada First Research Excellence Fund.
The numerical calculations were performed on computers provided by Calcul Qu\'ebec and Compute Canada.
We thank P. Lopes and P. Rinkel for helpful discussions.

\end{document}